\documentclass[final,leqno]{siamltex}

\def\d{\delta}

\usepackage{graphicx}
\usepackage{latexsym}
\usepackage{amssymb}
\usepackage{amsmath}
\usepackage{amsmath,empheq}
\usepackage{booktabs}
\usepackage[export]{adjustbox}
\usepackage{subfig}
\usepackage{enumerate}
\usepackage{multicol}
\usepackage{rotating}
\usepackage{pdflscape}
\DeclareMathSizes{12}{8}{7}{4}
\usepackage{nicefrac}
\usepackage{mathtools}
\usepackage{svg}
\usepackage{sidecap}
\usepackage{siunitx}
\usepackage{placeins}
\usepackage{comment}
\usepackage{sidecap}

\usepackage{xargs}                      

\usepackage[sc]{mathpazo}
\usepackage{algorithm}
\usepackage{algorithmicx}
\usepackage{algpseudocode}
\usepackage[sort,compress]{cite}
\usepackage{multirow}
\usepackage{makecell}
\usepackage{textcomp}
\usepackage{paralist}
\usepackage{cancel}
\usepackage{color, colortbl}
\usepackage{geometry}
\usepackage{longtable}
\usepackage{graphbox}

\usepackage[labelfont=bf,textfont=it,belowskip=0pt,aboveskip=5pt,tableposition=top]{caption}
\usepackage[colorlinks=true,
citecolor=blue,
filecolor=black,
linkcolor=blue,
urlcolor=blue]{hyperref}

\graphicspath{{figures/}{tables/}}

\algrenewcommand\alglinenumber[1]{
    {\sf\footnotesize}}
\algrenewcommand\algorithmicrequire{\textbf{Precondition:}}
\algrenewcommand\algorithmicensure{\textbf{Postcondition:}}

\newcounter{runidnum}
\newcommand{\runid}{\stepcounter{runidnum}\#\therunidnum}
\newcommand{\resetrunid}{\setcounter{runidnum}{0}}

\DeclareMathOperator*{\argmin}{arg\,min}

\newcommand{\figref}[1]{Fig.~\ref{#1}}
\newcommand{\tabref}[1]{Tab.~\ref{#1}}
\newcommand{\secref}[1]{\S\ref{#1}}

\newcommand{\sibia}{\textsf{SIBIA}}

\newcommand{\hazelhen}{\texttt{HazelHen}}

\newcommand{\nf}[2] {\nicefrac{#1}{#2}}

\newcolumntype{R}{>{\columncolor{gray!20}}r}
\newcolumntype{L}{>{\columncolor{gray!20}}l}

\newcommand{\Tfor}{\mathcal{T}}

\newcommand{\Rfor}{\mathcal{R}}

\newcommand{\diceB}{\ensuremath{\text{DICE}_B}} 
\newcommand{\diceHPB}{\ensuremath{\text{DICE}_{B_0}}} 
\newcommand{\diceT}{\ensuremath{\text{DICE}_T}} 
\newcommand{\elltwoB}{\ensuremath{\mu_{B,L^2}}} 
\newcommand{\elltwoHPB}{\ensuremath{\mu_{B_0,L^2}}} 
\newcommand{\elltwoT}{\ensuremath{\mu_{T,L^2}}} 
\newcommand{\relG}{\ensuremath{\|\vect{g}\|_{\text{rel}}}} 
\newcommand{\elltwoINIT}{\ensuremath{e_{c0,L^2}}} 
\newcommand{\errDiffInversion}{\ensuremath{e_{k_f}}} 
\newcommand{\rtIT}{T^{it}} 
\newcommand{\rtTUM}{T^{tu}_{inv}} 
\newcommand{\rtREG}{T^{reg}_{inv}} 

\newcommand{\mpPatientBrain}[1][]{\ifthenelse { \equal {#1} {} }
    { \vect{m}_D }   
    { \vect{m}_{D,#1} } }  

\newcommand{\mpWarpedAtlasBrain}[1][]{\ifthenelse { \equal {#1} {} }
    { \vect{m}_A^{(1,1)} }   
    { \vect{m}_{A,#1}^{(1,1)} }}   

\newcommand{\maHealthyAtlas}{\vect{m}_A}
\newcommand{\maHealthyAtlasT}{\vect{m}_P}
\newcommand{\maPatientBrain}{\vect{m}_D}
\newcommand{\maPatientPath}{c_D}
\newcommand{\maHealthyPatient}{\vect{m}_P(1)}
\newcommand{\maHealthyPatientit}[1]{\vect{m}_P(1)^{#1}}
\newcommand{\maSimBrain}{\vect{m}_P^{\prime}(1)}
\newcommand{\maSimPath}{c(1)}
\newcommand{\maSimPathit}[1]{c(1)^{#1}}

\newcommand{\maSimPathT}{c}
\newcommand{\initcond}{c(0)}

\newcommand{\Func}[3]{\ensuremath{\mathcal{#1}}_{#2}\left[#3\right]}    
\newcommand{\sFunc}[3]{\ensuremath{\mathcal{#1}}_{#2}[#3]}              
\newcommand{\D}[1]{\ensuremath{\mathcal{#1}}}                           

\newcommand{\ns}[1]{\ensuremath{\mathbf{#1}}}

\newcommand{\idiv}{\ensuremath{\mbox{div}\,}}
\newcommand{\igrad}{\ensuremath{\nabla}}
\newcommand{\ilap}{\rotatebox[origin=c]{180}{$\nabla$}}

\newcommand{\half}[1]{\frac{#1}{2}}

\renewcommand{\d}[1]{\mathop{}\!\mathrm{d}#1}
\newcommand{\dt}{\d{t}}

\newcommand{\p} {\partial}

\newcommand{\vect}[1]{\boldsymbol{#1}} 
\newcommand{\mat}[1]{\boldsymbol{#1}}  

\newcommand{\bipa}{\begin{inparaenum}[(\itshape i\upshape)]}
\newcommand{\eipa}{\end{inparaenum}}

\newcommand{\bipasub}{\begin{inparaenum}[(\itshape a\upshape)]}
\newcommand{\eipasub}{\end{inparaenum}}

\newcommand{\ipoint}[1]{\textit{\textbf{\color{darkgray}#1}}}

\newcommand{\iparagraph}[1]{~~{\color{black}\textit{\textbf{#1}}}.~~}
\newcommand{\iiparagraph}[1]{~~\textit{#1}.~~}

\newcommand{\ivar}[4]{\int_{#2}^{#3} #4 \, d#1}

\newcommand{\defeq}{\ensuremath{\mathrel{\mathop:}=}}

\definecolor{sred}{cmyk}{0.01,0.98,0,0.2} 
\reversemarginpar
\newcommand{\mmargin}[1]{{\marginpar{\em\tiny #1}}}\renewcommand{\mmargin}[1]{}

\usepackage[colorinlistoftodos,prependcaption,textsize=tiny]{todonotes}
\newcommandx{\unsure}[2][1=]{\todo[linecolor=red,backgroundcolor=red!25,bordercolor=red,#1]{#2}}
\newcommandx{\change}[2][1=]{\todo[linecolor=blue,backgroundcolor=blue!25,bordercolor=blue,#1]{#2}}
\newcommandx{\info}[2][1=]{\todo[linecolor=olive,backgroundcolor=olive!25,bordercolor=olive,#1]{#2}}
\newcommandx{\improvement}[2][1=]{\todo[linecolor=violet,backgroundcolor=violet!25,bordercolor=violet,#1]{#2}}
\newcommandx{\thiswillnotshow}[2][1=]{\todo[disable,#1]{#2}}

\title{Image-driven biophysical tumor growth  model calibration}

\author{Klaudius Scheufele\thanks{Institute for Parallel and Distributed Systems, Universit\"at
        Stuttgart, Universit\"atsstra{\ss}e 38, Stuttgart, Germany
        ({\tt klaudius.scheufele@ipvs.uni-stuttgart.de})}
        \and
        Shashank Subramanian \thanks{Oden Institute for Computational Engineering and Sciences, University of Texas at Austin,
      		201 E. 24th Street, Austin, Texas, USA
        ({\tt shashank@ices.utexas.edu})}
        \and
        Andreas Mang\thanks{Department of Mathematics, University of Houston, 3551 Cullen Blvd., Houston, Texas, USA
        ({\tt andreas@math.uh.edu})}
        \and
        George Biros\thanks{Oden Institute for Computational Engineering and Sciences, University of Texas at Austin,
      		201 E. 24th Street, Austin, Texas, USA
        ({\tt biros@ices.utexas.edu})}
        \and
        Miriam Mehl\thanks{Institute for Parallel and Distributed Systems, Universit\"at
        Stuttgart, Universit\"atsstra{\ss}e 38, Stuttgart, Germany
        ({\tt miriam.mehl@ipvs.uni-stuttgart.de})}
        }

\begin{document}

\maketitle

\begin{abstract}
We present a novel formulation for the calibration of a biophysical tumor growth model from a single-time snapshot,  multiparametric Magnetic Resonance Imaging (MRI) scan of a glioblastoma patient. Tumor growth models are typically nonlinear parabolic partial differential equations (PDEs). Thus, we have to generate a second snapshot to be able to extract significant information from a single patient snapshot. We create this two-snapshot scenario as follows. We use an atlas (an average of several scans of healthy individuals) as a substitute for an earlier, pretumor, MRI scan of the patient. Then, using the patient scan and the atlas, we combine image-registration algorithms and parameter estimation algorithms to achieve a better estimate of the healthy patient scan and the tumor growth  parameters that are consistent with the data.  Our scheme is based  on our recent work (\emph{Scheufele et al, ``Biophysically constrained diffeomorphic image registration, Tumor growth, Atlas registration, Adjoint-based methods, Parallel algorithms'', Computer Methods in Applied Mechanics and Engineering, 2018}), but apply a different and novel scheme where
the tumor growth simulation in contrast to the previous work is executed in the patient brain domain and not in the atlas domain yielding more meaningful patient-specific results.
As a basis, we use a PDE-constrained optimization framework.  We derive a modified Picard-iteration-type solution strategy in which we alternate between registration and tumor parameter estimation in a new way. In addition, we consider an $\ell_1$ sparsity constraint on the initial condition for the tumor and integrate it with the new joint inversion scheme. We solve the subproblems with a reduced-space, inexact Gauss-Newton-Krylov/quasi-Newton methods. We present results using real brain data with synthetic tumor data that show that the new scheme reconstructs the tumor parameters in a more accurate and reliable way compared to our earlier scheme.

\end{abstract}

\begin{keywords}
tumor progression inversion, biophysical model calibration, image registration, PDE constrained optimization, Picard iteration
\end{keywords}

\begin{AMS}
35K40, 49M15, 49M20, 65K10, 65N35, 65Y05, 92C50
\end{AMS}

\textit{\scriptsize Preprint submitted to the SIAM Journal on Scientific Computing (SISC).} \\

\pagestyle{myheadings}
\thispagestyle{plain}
\markboth{SCHEUFELE, SUBRAMANIAN, MANG, BIROS, MEHL}{Image-driven biophysical tumor growth  model calibration}

\section{Introduction}
\label{sec_scheuf:intro}

Glioblastoma multiforme (GBM) tumor is a  terminal primary brain cancer---the most aggressive one. Biophysical models are increasingly used to help the analysis of GBM MRI scans for epidemiological studies but also for assisting clinical decision making~\cite{Swanson:2002a,yankeelov-miga13}, survival estimation, diagnosis,  and preoperative and treatment planing~\cite{Mohamed:2006a,Jbabdi:2005,Gooya:2012a,Swanson:2008a}. The key step in integrating  biophysical models with clinical information is to calibrate them with patient MRI scans. After calibration,  we can either use the estimated parameters as biomarkers, or we can evolve the calibrated PDE to estimate short-term tumor infiltration. Tumor growth models are typically nonlinear reaction-diffusion PDEs and their calibration is challenging. Not only do we need to estimate  reaction and diffusion parameters, but also the tumor concentration, which is only partially (and implicitly) observed in the MRI scan. If we had an MRI scan of the patient without tumor,  we could solve an inverse problem for the initial condition, the reaction and diffusivity parameters of the tumor. Unfortunately, a \emph{``healthy patient''} MRI scan is rarely available. To resolve this conundrum, Davatzikos's group~\cite{Mohamed:2006a} pioneered the idea of using an MRI scan of another, \emph{healthy}, individual as a proxy for the tumor-free patient scan. In practice,  we use a standardized average brain of several individuals, also known as a \emph{statistical atlas}. But it turns out that naively using such an atlas would result in erroneous results. The second key idea addresses this by simultaneously using image registration to deform the atlas towards to the patient scan as far as possible. In this paper, we present a novel formulation of such a joint registration and inversion problem and a numerical solver scheme.

In the image registration problem, the goal is to estimate spatial point-correspondences between a template image $m_T$ (the atlas) and a reference image $m_R$ (the patient scan). To solve the registration problem, we use an optimal control formulation, in which we seek a (stationary) velocity field $\vect{v}$ (parameterizing a deformation map $\vect{y}$), such that the transported template image intensities match the intensities in the reference image, i.e., $m_T \circ \vect{y} \approx m_R$. In the tumor inversion problem, we want to estimate tumor-growth parameters (such as tumor origin, infiltration and proliferation rates) of our PDE model so that if we start growing the tumor at $t=0$ we match the partial tumor observations at $t=1$. (The astute reader is probably wondering how do we know the time horizon. We don't, but $t=1$ is related to a non-dimensional form of the tumor growth PDE.)  We simultaneously solve for both $\vect{v}$ and the tumor parameters.

\medskip
Let us try to explain the setting using a somewhat high-level notation. In our work we are not using the original MRI scan intensities, but assume that we have segmentation labels describing different brain tissue (white matter, gray matter etc.) distributions for both atlas and patient images.  Let $\vect{m}_A$ be the healthy atlas labels and $\vect{m}_D$ the patient labels (both vector functions in the unit cube in $\mathbb{R}^{3}$). Let $\Tfor$ and $\Rfor$, respectively, be abstract forward operators for a \ipoint{tumor simulation} and \ipoint{registration mapping} component.  $\Tfor( \vect{m}, \vect{p})$ takes a label image $\vect{m}$ without tumor and tumor model parameters $\vect{p}$ and creates a labeled image that now has tumor labels in addition to the original healthy tissue labels (e.g., gray matter, white matter, etc.). $\Rfor(\vect{m}, \vect{v})$ takes a label image $\vect{m}$ and a velocity $\vect{v}$ that parameterizes the deformation and creates a deformed image.

\begin{figure}
\begin{center}
\includegraphics[width=.7\textwidth]{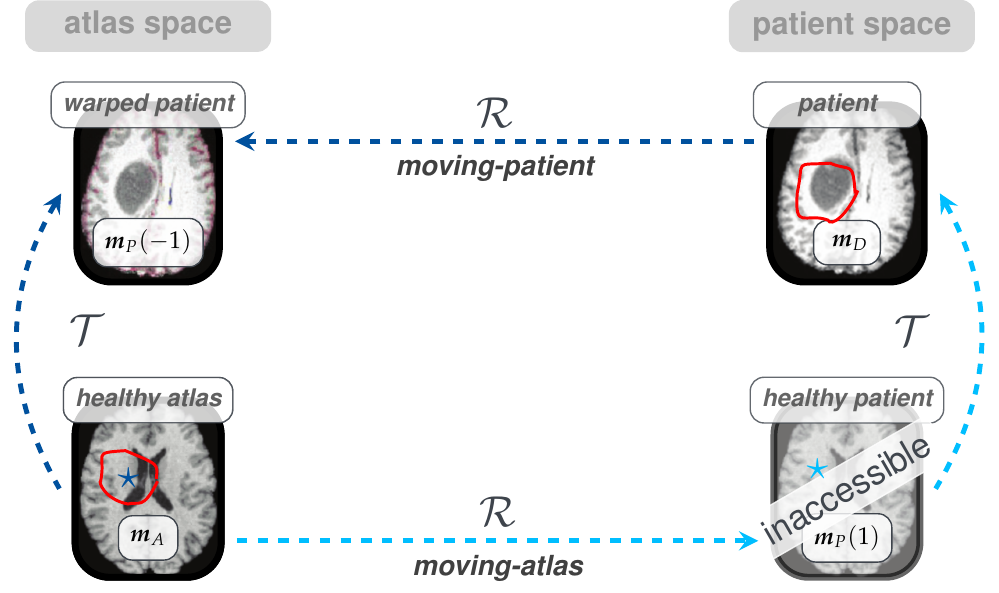}
\end{center}
 \caption{Schematic view of the \textit{Moving Patient} and the (new) \textit{Moving Atlas} scheme.
 The \textit{Moving Patient} coupling-scheme performs the tumor simulation (operator $\Tfor$)  (and parameter estimation) in a healthy atlas brain and warps the patient anatomy and tumor labels towards the atlas space (operator $\Rfor$)~\cite{Scheufele:2018,Scheufele:2019}.
 For the new \textit{Moving Atlas} coupling-scheme, we first compute an approximation of the actual healthy patient brain by warping the atlas anatomy labels towards the patient (operator $\Rfor$) and perform the tumor simulation (operator $\Tfor$) (and parameter estimation) in the resulting approximation of the healthy brain geometry. This is critical to obtain patient specific tumor parameter estimates (of, e.g., tumor origin, tumor cell migration rate, tumor cell proliferation rate) that may be of (future) clinical relevance in decision making and treatment.
 Brain images modified from~\cite{Gooya:2012a}.
 \label{fig:patient_vs_atlas}}
\end{figure}

Then, we can summarize the general idea of our previous \textit{Moving-Patient} and the new \textit{Moving-Atlas} scheme by defining two optimization problems:
\begin{equation}\label{e:ma-mp-objective}
\begin{aligned}
   &\mbox{\textit{Moving-Patient (MP):}} & &\mbox{min}_{\vect{p},\vect{v}} \| \Rfor\left( \Tfor(\vect{m}_A, \vect{p}), \vect{v} \right) - \vect{m}_D \|, \\
   &\mbox{\textit{Moving-Atlas (MA):}} & &\mbox{min}_{\vect{p},\vect{v}} \| \Tfor\left( \Rfor(\vect{m}_A, \vect{v}), \vect{p} \right) - \vect{m}_D \|.
\end{aligned}
\end{equation}
Notice that the main difference is switching  the order of the registration and the tumor operators; see \figref{fig:patient_vs_atlas} for an illustration. This seemingly simple change has significant impact in the solution of this problem. In~\cite{Scheufele:2018,Scheufele:2019}, we used the \emph{Moving-Patient} scheme and only inverted for the tumor initial condition. Note that in \textit{MP}, to evaluate the objective function,
we first grow a tumor in the atlas and then deform the patient image to match the resulting tumor-bearing atlas.
As long as the tumor-bearing atlas is topologically similar to the tumor-bearing patient brain, image registration can always
yield good tumor reconstruction, even with completely wrong tumor parameters. Thus this approach enables efficient
non-diffeomorphic registration between an atlas and a tumor-bearing patient, but is limited in terms of meaningful tumor inversion.

In this paper,  we propose the \textit{Moving Atlas} scheme. To evaluate the objective function given, we first deform the healthy atlas (using $\vect{v}$) so that it matches the  patient. Then we grow a tumor in the transformed atlas. It turns out that the new formulation is more appropriate for the biophysical modeling since the tumor growth takes place in images that resemble the actual patient.

\medskip\noindent
\iparagraph{Contributions}
The main contributions of this paper are as follows:
\begin{enumerate}[(i)]
  \item We present a new optimization problem formulation (\textit{Moving Atlas}) for tumor growth model calibration based
  on patient individual single-snapshot data.
  \item The new formulation prevents fitting of patient input data to a possibly poor tumor reconstruction. The computed biophysical parameters ``live'' in the patient space, i.e., the tumor parameters are estimated assuming tumor growth in (an approximation of) the healthy patient brain (as opposed to the atlas brain for the former scheme). This renders the new scheme better for biophysical inversion.\footnote{The presented results are to be seen as a proof of concept for the developed methodology. To enable predictive capabilities, we require a more complex tumor model.}
  \item We derive a Picard-iteration-type solution scheme that alternates between the image registration and the inverse tumor-growth problem.\footnote{A modified objective function for the registration sub-component allows to fulfill the strongly coupled first order optimality conditions of the joint optimization problem formulation}
  \item We enhance the tumor inversion component. The tumor solver used in~\cite{Scheufele:2018} inverted for the tumor initial condition only, penalized with an $\ell_2$ constraint. In this paper, we use an $\ell_1$ constraint (which restricts the tumor to more plausible initial condition), and also invert for the diffusion parameter in the tumor growth PDE (modeling tumor infiltration).
   \item We conduct numerical experiments to evaluate the nem scheme using synthetically grown tumor data from real clinical brain imaging data. We demonstrate that the \textit{Moving Atlas} scheme yields better results in terms of the accuracy and robustness of tumor parameter reconstruction.\footnote{For all considered test cases the tumor growth ground truth parameters are known, to allow for comparison of the respective reconstruction accuracy.} For a synthetic example, we showcase malfunction\footnote{The patient input data is fitted to a poor reconstruction of the tumor in the atlas (due to large anatomical differences); consequently, the target data is corrupted in the further course of the inversion.} of the \textit{Moving Patient} scheme, whereas the new \textit{Moving Atlas} scheme scheme yields good results.
\end{enumerate}

\medskip\noindent
\iparagraph{Limitations}
The reaction-diffusion tumor-growth model we consider is purely phenomenological. Important biophysical, bio-mechanical and biochemical effects and phenomena such as mitosis, apoptosis, chemotaxis, deformation of brain parenchyma (mass-effect), and the modeling of edema, necrosis, and angiogenesis are neglected. It is, however, the most common model used for clinical analysis~\cite{Swanson:2000a,Swanson:2008a,Clatz:2005a,Harpold:2007a,Konukoglu:2010a,Hogea:2008a}. The effect of tumor growth induced deformation of brain parenchyma is important and not accounted for in our current model. This is ongoing work. Note that the main work here lies in the enhancement of the tumor component. Integrating it in the coupled formaulation and the Picard iteration is straight-forward due to our modular approach. The inverse tumor-growth problem with inversion for all growth parameters is ill-posed. We have to introduce additional prior information. Currently, if we want to invert for the proliferation rate $\rho$ and the diffusion rate, we run the solver for multiple values of $\rho$ and compare the quality of reconstruction results. We do not account for uncertainties in input imaging data, employed algorithms and solution approaches, and model parameters (such as tumor initial condition and characteristic net cell migration into surrounding tissue). A Bayesian framework is subject to future work; we currently invert for the maximum a posteriori estimate of the parameters.
Our approach does not directly process MRI data but requires prior segmentation of the imaging data. Note that reliably integrating models with MRI is an open problem, since one needs to estimate cell-density from data or link model output to intensities. Lastly, we emphasize that within this work we present a proof-of-concept for a new, promising methodology; future work will include its application to real clinical data.

\medskip\noindent
\iparagraph{Related Work}
One important application of biophysical models is to enable non-diffeomorphic image registration. Classical image registration~\cite{Modersitzki:2004a,Sotiras:2013a} assumes that the input images (template and reference) are topologically equivalent. But registering a healthy atlas (no tumor) to a tumor-bearing patient image violates this assumption.  We refer to this problem as the normal-to-abnormal registration. We refer to our previous work in~\cite{Scheufele:2018}  for an extensive review in non-diffeomorphic  image registration.

Here, we are interested in fitting models to patient-individual imaging data. Several groups have tackled this problem using derivative-free optimization approaches~\cite{Chen:2012b,Knopoff:2017a,Konukoglu:2010a,Mang:2014a,Mang:2012a,Mi:2014a,Wong:2015a,Wong:2017a}, or address the parameter estimation problem within a derivative-free Bayesian framework~\cite{HawkinsDaarud:2013b,Le:2015a,Le:2017a,Lima:2016a,Lima2017a,Menze:2011a,Oden:2013}. There has been only limited work in the direction of PDE-constrained optimization for model-based image analysis~\cite{Colin:2014a,Hogea:2008a,Gholami:2016a,Knopoff:2017a,Knopoff:2013a,Liu:2014a,Mang:2014a,Mang:2012b,Quiroga:2015a}. Most of  these approaches rely on longitudinal patient data. In particular, they require knowledge of the healthy patient brain before tumor occurrence. To allow for model-inversion based on single-snapshot data only, deformable inter-subject registration becomes necessary to artificially generate a second snap-shot in time. Such a combination of registration and biophysical inversion has been previously targeted in~\cite{Gooya:2012a,Hogea:2008a,Mohamed:2006a,Zacharaki:2008a,Zacharaki:2008b,Zacharaki:2009a}. The work in~\cite{Mohamed:2006a,Zacharaki:2008a,Zacharaki:2008b,Zacharaki:2009a}, however, uses a purely mechanical model for tumor progression and falls short in providing information about progression and infiltration of cancerous cells into surrounding healthy tissue.
In~\cite{Bakas:2015a,Kwon:2014a,Hogea:2008a,Gooya:2012a}, the authors propose a framework for joint segmentation, registration and tumor modeling, which is very similar to our approach. In our previous work~\cite{Scheufele:2018}, we significantly reduce the time-to-solution by employing second-order derivative information in conjunction with highly scalable and efficient numerics, improved algorithms and powerful preconditioners. We eliminate the need for manual tumor seeding, and improve inter-subject registration performance.

\medskip\noindent
\iparagraph{Outline}
The outline of the paper is as follows: In~\secref{sec_scheuf:moving_atlas} we present the mathematical formulation for the
\textit{Moving Atlas} joint registration and biophysical inversion problem. We present the PDE-constrained optimization problem and derive the first order optimality conditions in~\secref{sec_scheuf:ma_formulation} and propose a Picard-iteration-type solution strategy, outlined in~\secref{sec_scheuf:picard}. Details on continuation schemes, adaptivity and convergence criteria are given in~\secref{sec_scheuf:adaptivity-conv-crit}. More details on the two main sub-components, diffeomorphic medical image registration and biophysical tumor-growth inversion are summarized in~\secref{sec_scheuf:components}. In~\secref{sec_scheuf:results}, we perform numerical experiments to analyze our scheme, and compare it to the \textit{Moving Patient} scheme~\cite{Scheufele:2018}.

\section{Formulation and Picard Iteration for the Moving Atlas Coupled Problem}
\label{sec_scheuf:moving_atlas}
To apply tumor inversion and image registration in a joint approach, we consider an optimal control formulation, which results in a PDE-constrained, nonlinear optimization problem. In this section, we present the new \textit{Moving Atlas} formulation along with an iterative fixed-point coupling scheme. This scheme allows us to solve the joint optimization problem based on two separate components: a tumor-growth inversion solver and a modified diffeomorphic image registration component. Before presenting the formulation and the iterative solver, we shortly introduce the notation we are going to use throughout the paper.

\begin{table}
\small\centering\setlength\tabcolsep{4pt}
\caption{
	Notation for the \ipoint{Moving Atlas} joint tumor inversion and image registration formulation:
	$\vect{m}$ denotes the vector of anatomy labels (probability maps for white matter (WM), gray matter (GM) and cerebrospinal fluid (CSF)) defining the brain geometry, and $c$ denotes a probability map for tumor (TU). Subscripts $A$ and $P$ indicate variables in the atlas and patient space, respectively. Anatomy labels and tumor label with subscript $D$ denote the patient input imaging data (after prepossessing). The fields evolve in time and space; anatomy labels $\vect{m}$ for brain tissue evolve along a pseudo-time $t_R \in [0,1]$ associated to registration, and tumor labels evolve along the (normalized) time $t_T \in [0,1]$ of tumor growth. The integration of a simulated tumor map $c(1)$ into healthy brain tissue $\vect{m}$ is indicated by $\vect{m}^{\prime}$ and modeled via the formula $\vect{m}^{\prime} = \vect{m} (1-\maSimPath)$.  To judge the proximity of the predicted state of our mathematical model to the observed data $\maPatientBrain$, and $\maPatientPath$, we define the $\ell_2$-distance measures $\D{D}_c$ and $\D{D}_{\vect{m}}$.
\label{tab:notation}}
\begin{tabular}{lll}
\toprule
 \textit{healthy atlas brain (input)}                     & $\maHealthyAtlas$ & \multirow{9}{*}{\includegraphics[width=5cm]{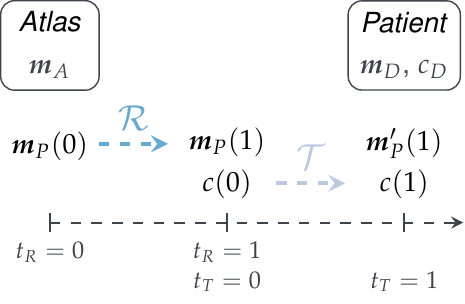}} \\
 \textit{patient brain with tumor (input; target data)}   & $\maPatientBrain$ \\
 \textit{patient tumor (input; target data)}              & $\maPatientPath$ \\
 \textit{healthy patient brain}                           & $\maHealthyPatient$ \\
 \textit{tumor initial condition}                         & $\initcond$ \\
 \textit{simulated tumor}                                 & $\maSimPath$ \\
 \textit{approximated patient brain}                      & $\maSimBrain$ \\
 \textit{$\ell_2$-misfit tumor}                           & $\Func{D}{c}{c_1,c_2} \defeq \half{1} \|c_1 - c_2\|^2_{L^2(\Omega)}$ \\
 \textit{$\ell_2$-misfit anatomy labels}                  & $\Func{D}{\vect{m}}{\vect{m}_1,\vect{m}_2} \defeq \half{1} \|\vect{m}_1 - \vect{m}_2 \|^2_{L^2(\Omega)^3}$ \\
\bottomrule
\end{tabular}
\end{table}

\medskip\noindent
\iparagraph{Notation}
$m_X(\vect{x},t) \in [0,1]$, $(\vect{x},t) \in \Omega \times [0,1]$, with $X \in \{WM, GM ,CSF \}$ represents probability maps of different brain tissue types, namely, white matter (WM), gray matter (GM), and cerebrospinal fluid (CSF; which includes ventricles). The probability maps for each tissue label are computed from (segmented) MR imaging data in a pre-processing step. In the following, we refer to $m_X(\vect{x},t)$ as anatomy labels.  $\Omega = [0,2\pi]^3$ denotes the normalized spatial domain. We gather these probability maps in  a space-time vector field
\begin{equation}
\vect{m}(\vect{x},t) =
\left( m_i(\vect{x},t) \right)_{i = WM, GM, CSF} \in\ns{R}^3 \;
\label{e:vec_m}
\end{equation}
The normalized tumor cell concentration $c \colon \Omega_B \times [0,1] \rightarrow [0,1], (\vect{x}, t) \mapsto c(\vect{x}, t)$ (tumor map) is given as a fourth field and interpreted as probability map for cancerous tissue. $\Omega_B \subset \Omega = [0,2\pi)^3$ is embedded\footnote{We use a penalty approach to discretize the tumor equations in $\Omega$ and approximate the Neumann boundary conditions on $\partial\Omega_B$.} in the simulation domain $\Omega$ and denotes the domain occupied by brain tissue; it is bounded by the skull. All anatomy and tumor labels evolve in space and time. To simplify notation, we introduce the space-time domain $U \defeq \Omega \times (0,1]$, and $\bar{U} \defeq \Omega \times [0,1)$. We omit the spatial dependency, and indicate temporal evolution by the pseudo time $t_R\in [0,1]$ for the advection problem of the registration, and the normalized time $t_T\in[0,1]$ for tumor growth, cf.\;\tabref{tab:notation}. To judge the proximity of the predicted state of our mathematical model to the observed data $\maPatientBrain$, $\maPatientPath$, we define the $\ell_2$-distance measures $\D{D}_c$ and $\D{D}_{\vect{m}}$ (see~\tabref{tab:notation}), which drive the optimization process.

\subsection{Moving Atlas Coupled Formulation}
\label{sec_scheuf:ma_formulation}

The \textit{Moving Atlas} coupled problem formulation is given by the minimization problem
\begin{subequations}  \label{eq:ma:global_opt}
\begin{gather}\label{e:ma:global_objective}
\min_{\vect{v},\vect{p},k_f,w}~  \mathcal{D}_c[\maSimPath,\maPatientPath]   +
\mathcal{D}_{\vect{m}}[\maHealthyPatient (1 - \maSimPath),\maPatientBrain] +
\D{S}[\vect{p},\vect{v},w] \notag
\end{gather}
\noindent subject to
\begin{flalign}
&\mbox{~~~~~~~$\Tfor\colon$ \emph{(tumor forward op.)}} & \p_t \maSimPathT - \idiv \mat{k}(\maHealthyPatient) \igrad \maSimPathT - f( \maSimPathT, \rho(\maHealthyPatient) ) &=  0  &&\mbox{in}~U, \label{e:tumor:fwd} \\
 & &\Phi(\maHealthyPatient) \vect{p} &= \initcond  &&\mbox{in}~\Omega, \label{e:tumor:fwd-init} \\
&\mbox{~~~~~~~$\Rfor\colon$ \emph{(registration advection op.)}} & \partial_t \maHealthyAtlasT + \nabla \maHealthyAtlasT \cdot \vect{v} &= \vect{0}  &&\mbox{in}~\bar{U},
             \label{e:reg:fwd} \\
 & &\maHealthyAtlas &= \maHealthyAtlasT(0) &&\mbox{in}~\Omega,
              \label{e:reg:fwd-init} \\
 & &\idiv \vect{v} &= w  & &\mbox{in}~U,
\label{eq:reg:divfree}
\end{flalign}
\end{subequations}
That is, we seek parameters and space time fields $(\vect{v}, \vect{p}, k_f, w)$, such that the predicted state of our model becomes similar to the observed data. The distance between data and predicted state is quantified by the two $\ell_2$-distance measures $\sFunc{D}{c}{\maSimPath,\maPatientPath}$ and $\sFunc{D}{\vect{m}}{\maHealthyPatient (1 - \maSimPath), \maPatientBrain}$.
$\sFunc{D}{c}{\maSimPath,\maPatientPath}$ measures the discrepancy between the simulated tumor $\maSimPath$ computed from~\eqref{e:tumor:fwd}--\eqref{e:tumor:fwd-init} in the \ipoint{approximated patient space} and the input tumor data $\maPatientPath$.
$\sFunc{D}{\vect{m}}{\maHealthyPatient (1 - \maSimPath), \maPatientBrain}$ quantifies the discrepancy between the patient's
anatomy labels and the \ipoint{warped-to-patient} atlas anatomy labels in regions not affected by the tumor\footnote{The simulated pathology is embedded into the advected atlas anatomy labels when caculating the registration misfit. This can be seen as a ``masking'' of the tumor region, i.e., the registration does not alter the tumor concentration nor the discrepancy measure $\D{D}_{c}$.} computed from~\eqref{e:reg:fwd}--\eqref{eq:reg:divfree}).
$\D{S}[\vect{p},\vect{v},w]  \defeq \beta_{\vect{p}} \D{S}_{\vect{p}}[\vect{p}] + \beta_{\vect{v}} \D{S}_{\vect{v}}[\vect{v}] + \beta_w \D{S}_w[w]$ is a regularization term.
%

To simulate \ipoint{tumor-growth} over time, we use a diffusion-reaction model with diffusion tensor $\mat{k}$ and a logistic reaction term $f(c, \rho) = \rho c (1-c)$ with reaction rate $\rho$.
For the tumor initial conditions, we use a parametrization $\Phi\vect{p}$, i.e., $\initcond$ lives in an $n_{\vect{p}}$-dimensional space spanned by Gaussian basis functions ($\vect{p} \in\ns{R}^{n_{\vect{p}}}$, $\Phi \vect{p} := \sum_{i=1}^{n_{\vect{p}}} \Phi_i p_i$). Within CSF, the Gaussians are set to zero to prevent spurious diffusion of cancerous cells into CSF. We model the \ipoint{diffusion} and \ipoint{reaction} coefficient as
\begin{equation}\label{eq:coeff_tumor}
  \mat{k}(\vect{m}) \defeq  k_f\, \sum_{i=1}^3 k_i m_i\mat{I} = \overline{k}  \vect{m}\mat{I} ~~~~\mbox{and}~~~~
  \rho(\vect{m})    \defeq \rho_f  \sum_{i=1}^3 \rho_i m_i = \overline{\rho} \vect{m}
\end{equation}
with $\overline{k} = ( k_1 k_f, k_2  k_f, k_3  k_f )^T$ and $\overline{\rho} = ( \rho_1 \rho_f, \rho_2 \rho_f, \rho_3 \rho_f )^T$,  i.e., both diffusion and reaction parameters vary in space depending on the
value of the anatomy labels $\vect{m}$ describing the brain geometry.
For simplicity, we only consider isotropic diffusion\footnote{Note, that our approach can be generalized in a straight-forward way to general diffusion tensors}. To regularize the tumor growth initial condition parametrization, we use combinations of $\ell_1$- and $\ell_2$-regularization, i.e.,
\begin{equation}
\D{S}^1_{\vect{p}}[\vect{p}] = \frac{1}{2} \|\vect{p}\|_{L^1(\Omega)}^2 \, \mbox{ and } \,
\D{S}^2_{\vect{p}}[\vect{p}] = \frac{1}{2} \|\vect{p}\|_{\textbf{W}(\Omega)}^2 := \sum_{i=1}^{p_n} w_i \cdot p_i^2,
\label{eq_scheuf:tumor_reg}
\end{equation}
where $\beta_{\vect{p}} > 0$ is the regularization parameter and $\| \cdot \|_{\textbf{W}(\Omega)}$ is a weighted $\ell_2$-norm. For more details on how we switch between the two penalty terms in the solver algorithm, see \secref{sec_scheuf:adaptivity-conv-crit}.

For \ipoint{image registration}, we use an advection model with a stationary velocity field $\vect{v}$. The image registration
velocity is regularized via the $H^1$ Sobolev semi-norm:
\begin{equation}
\D{S}_{\vect{v}}[\vect{v}] = \frac{1}{2} \| \vect{v} \|_{H^1(\Omega)}^2 = \frac{1}{2} \int_{\Omega} \sum_{i=1}^3| \nabla v^i(\vect{x})|^2\d\Omega. \label{eq_scheuf:registr_reg}
\end{equation}$\D{S}_w[w] = \| w \|_{L^2(\Omega)}$ is a regularization term for $w = \idiv{\vect{v}}$, which allows us to control volume change.
We use periodic boundary conditions on $\partial\Omega$.

\medskip\noindent
\iparagraph{Optimality Conditions} As a general optimization approach, we use the method of Lagrange multipliers. Taking variations yields the strong form of the \ipoint{first order optimality conditions}:
%
\begin{subequations}\label{e:ma:first_order_opt_global}
\begin{align}
 & \mbox{\textit{tum., state}} %
&  \p_t \maSimPathT - \idiv \mat{k}(\maHealthyPatient) \igrad \maSimPathT - f( \maSimPathT, \rho(\maHealthyPatient) ) &=  0  & &\mbox{in}~U, \label{e:ma:glob:opt-cond:tumor:state}  \\
&
&  \initcond - \Phi(\maHealthyPatient) \vect{p} &= 0 & &\mbox{in}~\Omega,           \label{e:ma:glob:opt-cond:tumor:state-init}  \\
& \mbox{\textit{tum., adj.} }
&             -\partial_t \alpha - \idiv \mat{k}(\maHealthyPatient)\igrad \alpha  + \partial_{\maSimPathT} f^\star(\maSimPathT, \rho(\maHealthyPatient)) \alpha &= 0 & &\mbox{in}~\bar{U},
             \label{e:ma:glob:opt-cond:tumor:adj} \\
&
&             \maPatientPath - \maSimPath + (\maHealthyPatient)^T(\maHealthyPatient(1- \maSimPath) - \maPatientBrain) - \alpha(1) &= 0 & &\mbox{in}~\Omega,
             \label{e:ma:glob:opt-cond:tumor:adj-final} \\
& \mbox{\textit{reg., state} }
& \partial_t \maHealthyAtlasT + \nabla \maHealthyAtlasT \vect{v} &= \vect{0}  &&\mbox{in}~\bar{U},            \label{e:ma:glob:opt-cond:reg:state} \\
&
& \maHealthyAtlasT(0) -  \maHealthyAtlas &= \vect{0} &&\mbox{in}~\Omega,             \label{e:ma:glob:opt-cond:reg:state-init} \\
& \mbox{\textit{reg., adj.} }
&              -\partial_t \vect{\lambda} -\idiv (\vect{\lambda} \otimes \vect{v})  &= \vect{0} &&\mbox{in}~\bar{U},
              \label{e:ma:glob:opt-cond:reg:adj} \\
&
&             (\maSimPath -1)(\maHealthyPatient (1 - \maSimPath) - \maPatientBrain) + \partial_{\maHealthyPatient} \Phi(\vect{p}) \alpha(0) -
						 &  && \label{e:ma:glob:opt-cond:reg:adj-final}\\
             & & & \llap{$\displaystyle  \ivar{t}{0}{1}{(\igrad \maSimPathT)^T \igrad \alpha \, \partial_{\maHealthyPatient}\mat{k}^\star(\maHealthyPatient)
             + \partial_{\maHealthyPatient} f^\star(\maSimPathT, \rho(\maHealthyPatient)) \alpha}
             - \vect{\lambda}(1)$} = \vect{0}   & &\mbox{in}~\Omega,
              \notag  \\
& \mbox{\textit{tum., inv.} }
&               \beta_{\vect{p}} \igrad_{\vect{p}} \D{S}_{\vect{p}}[\vect{p}] - \Phi^T \alpha(0) &= \vect{0} & &\mbox{in}~\Omega,
               \label{e:ma:glob:opt-cond:tumor:inv} \\
&
&\int_0^1 \int_{\Omega} \maHealthyPatient \left( (\igrad \maSimPathT)^T \igrad\alpha \right) \, dx\,dt & = \vect{0} &&\mbox{in}~\Omega,
             \label{e:ma:glob:opt-cond:tumor:inv_diff} \\
& \mbox{\textit{reg., inv.} }
&               \beta_{\vect{v}} \igrad_{\vect{v}} \D{S}_{\vect{v}}[\vect{v}] + \D{K}[\int_0^1 (\igrad \maHealthyAtlasT)^T \vect{\lambda}\, \dt] &= \vect{0}   & &\mbox{in}~\Omega,
               \label{e:ma:glob:opt-cond:reg:inv}
\end{align}
\end{subequations}
with adjoint variables $\alpha$, $\vect{\lambda}$, $\nu$ associated to the state variables $\maSimPath$, $\maHealthyAtlasT$, and
$\maHealthyPatient$. We eliminate $w$ from the optimization problem and use the resulting operator $\D{K}$ in \eqref{e:ma:glob:opt-cond:reg:inv}. For $\idiv \vect{v} = w=0$, $\D{K}$ would be the Leary projection $\D{K}(\vect{u}) \defeq \vect{u} + \igrad \ilap^{-1}\idiv \vect{u}$; for a non-zero $w$, the projection operator becomes slightly more involved; we refer to~\cite{Mang:2015a,Mang:2016a} for additional details.
The gradients of the diffusion and reaction terms in \eqref{e:ma:glob:opt-cond:reg:adj-final} with respect to $\maHealthyPatient$ can be derived from \eqref{eq:coeff_tumor}, and are given by:
\begin{align}
& & \partial_{\maHealthyPatient} \mat{k} &=  \bar{k} = k_f(k_1,k_2,0)
& \mbox{and} &
& \partial_{\maHealthyPatient} \mat{f} &= \overline{\rho}\maHealthyPatient(1-2 \maSimPath).  & \label{e:c7:ma:diff-and-reac-coeff-gradients}
\end{align}

\medskip \noindent
\iparagraph{Discussion}
We summarize the advantages of the moving atlas scheme over the previously used moving patient scheme \cite{Scheufele:2018}:
\bipa
\item image registration operates on the brain anatomy only such that the tumor solver is forced to actually produce a biophysically more
meaningful\footnote{Assuming a biophysical tumor progression model which accurately reflects the physiological processes; the model used here does not fulfill this assumption.} tumor that is similar to the observed patient tumor.
\item Tumor growth parameters, i.e., $\vect{p}$ and $k_f$, are estimated directly in an approximation of the patient brain such that their quality does not strongly deteriorate with differences between the atlas and the patient brain.
\eipa
This comes at a price, for which we account for by employing suitable solver strategies presented in \secref{sec_scheuf:components}:
\bipasub
\item the decomposition of the moving atlas formulation into tumor and registration components requires modifications of the registration formulation as shown in \secref{sec_scheuf:picard},
\item exploiting the full potential of the moving atlas scheme requires sparse localization of the tumor initial condition, encouraged by an $\ell_1$-regularization in the tumor component as outlined in~\secref{sec_scheuf:tumor_reg} (cf. also~\cite{Subramanian19aL1,Scheufele:2019}).
\eipasub

\subsection{Picard Iteration Scheme}
\label{sec_scheuf:picard}

We solve the optimization problem by iterating over the registration velocity $\vect{v}$, the tumor growth initial guess parametrization $\vect{p}$ and (if switched on) the diffusion coefficient $k_f$. We use a superscript $j$ to mark variables associated with the $j$th Picard iteration.
Each Picard iteration is decomposed two into sub-steps:
\begin{enumerate}[(1)]
  \item Solve \ipoint{inverse tumor problem}, i.e., given $\vect{v}^j$ solve for $(\vect{p}^{j+1}, k_f^{j+1})$ such that
				\begin{equation}\label{e:objective_tu}
				(\vect{p}^{j+1}, k_f^{j+1}) = \argmin_{\vect{p}, k_f}~
           \underbrace{\mathcal{D}_c[\maSimPath,\maPatientPath]   + \mathcal{D}_{\vect{m}}[\maHealthyPatientit{j} (1 - \maSimPath),\maPatientBrain]
					 + \beta_{\vect{p}} \D{S}_{\vect{p}}[\vect{p}]}_{\mbox{$=: \mathcal{J}_T$}}
        \end{equation}
        \ipoint{subject to} the tumor forward problem \eqref{e:tumor:fwd} and \eqref{e:tumor:fwd-init} for the healthy patient anatomy $\maHealthyPatient = \maHealthyPatientit{j}$ kept fixed (result of the previous Picard iteration or initial guess). This means, we fulfill the first order optimality conditions \eqref{e:ma:glob:opt-cond:tumor:state}--\eqref{e:ma:glob:opt-cond:tumor:adj-final}, \eqref{e:ma:glob:opt-cond:tumor:inv} and \eqref{e:ma:glob:opt-cond:tumor:inv_diff} for $\maHealthyPatient = \maHealthyPatientit{j}$. Aside from new iterates for $\vect{p}$ and $k_f$, this step yields a new simulated tumor $\maSimPathit{j+1}$ and tumor adjoint $\alpha^{j+1}$.

  \item Solve \ipoint{modified registration problem}, i.e., given $(\vect{p}^{j+1}, k_f^{j+1})$, solve for $\vect{v}^{j+1}$ such that
				\begin{equation}\label{e:objective_reg}
				( \vect{v}^{j+1}, w^{j+1}) = \argmin_{\vect{v}, w}~
            \mathcal{D}_{\vect{m}}[\maHealthyPatient (1 - \maSimPathit{j+1}),\maPatientBrain] +  \int_{\Omega} \vect{q}^T \maHealthyPatient \, \d{\vect{x}} +
            \beta_{\vect{v}}\D{S}_{\vect{v}}[\vect{v}] + \beta_{w}\D{S}_{w}[w]
        \end{equation}
        \ipoint{subject to} the image advection problem \eqref{e:reg:fwd}--\eqref{eq:reg:divfree}. $\vect{q} = (q_1, q_2, q_3)^T$ is defined as
        $$\vect{q}(\vect{x}) = \int_{0}^{1}{\overline{k}((\igrad \maSimPathit{j+1})^T \igrad \alpha^{j+1}) + \overline{\rho}\maSimPathit{j+1}(1-\maSimPathit{j+1})\alpha^{j+1}}\,\d{t}.$$
  \item iterate until convergence.
\end{enumerate}

\medskip
The particular modification of~\eqref{e:objective_reg} by introducing the term $\int_{\Omega} \vect{q}^T \maHealthyPatient \, \d{\vect{x}}$ ensures that the registration problem reproduces~\eqref{e:ma:glob:opt-cond:reg:state} -- \eqref{e:ma:glob:opt-cond:reg:adj-final} and \eqref{e:ma:first_order_opt_global} as first order optimality conditions. Note that we freeze the variables $\maSimPath$, and $\alpha$ from the previous iteration, i.e., neglect the indirect impact of the registration velocity $\vect{v}$ on $\maSimPath$ and $\alpha$ via changes in $\maHealthyPatient$. Besides new iterates for $\vect{v}$, the registration step also provides a new approximation $\maHealthyPatientit{j+1}$ of the healthy patient brain.
The converged solution of the respective Picard or fixed-point iteration fulfills all first order optimality conditions of \eqref{eq:ma:global_opt} with the exception of~\eqref{e:ma:glob:opt-cond:reg:adj-final}. For the latter, the term $\partial_{\maHealthyPatient} \Phi(\vect{p}) \alpha(0)$ is neglected and not enforced explicitly. We found experimentally, that our scheme works well and produces satisfying results despite the simplification.

\subsection{Adaptivity and Convergence Criteria}
\label{sec_scheuf:adaptivity-conv-crit}
In our Picard scheme, we use several strategies to control convergence and to adapt the tumor and image registration regularization over the Picard iterations.

For the tumor solver, the regularization term $\beta_{\vect{p}} \D{S}_{\vect{p}}[\vect{p}]$ is used to attain better conditioning, but also to encourage sparse localization of the initial condition. Thus, we use $\D{S}_{\vect{p}}[\vect{p}] = \nf{1}{2} \|\vect{p}\|_{L^1(\Omega)}^2$ to compute a sparse solution of the initial condition and switch to a weighted $\ell_2$-regularization $\D{S}_{\vect{p}}[\vect{p}] = \|\vect{p}\|^2_{\mat{W}(\Omega)} = \nf{1}{2} \sum_{i=1}^{p_n} w_i \cdot p_i^2$ in a second phase of the tumor inversion solver, where the weights $w_i$ are calculated according to
\[
w_i = \left\{ \begin{array}{ll} w^{\text{large}} \, \mbox{ if } \, | p_i^{\ell_1} | < \tau_S \|\vect{p}\|_{L^\infty(\Omega)} \\
                                   w^{\text{small}} \, \mbox{ otherwise.} \end{array} \right.
\]
Here, $\tau_S$ is a user defined tolerance. The $\ell_2$-phase uses $\vect{p}^{\ell_1}$ achieved in the $\ell_1$-phase as initial approximation. In the $\ell_1$-phase we invert for $\vect{p}$ only and keep the diffusivity $k_f$ (estimated in the $\ell_2$-phase) fixed;\footnote{We found experimentally, that diffusion inversion in the $\ell_1$-phase has vanishing impact on the proposed sparsity of the solution, and the $\ell_2$ estimation for $k_f$ is better.} initially, we start with $k_f=0$. For more details on numerics of the tumor inversion solver, we refer to~\cite{Scheufele:2018}.

For the registration solver, convergence and robustness are improved by regularization. We perform a parameter continuation. That is, we start with a large value $\beta_{\vect{v}}^{\text{init}}$ and subsequently reduce $\beta_{\vect{v}}$ by a factor of ten after each Picard iteration until we reach the prescribed value $\beta_{\vect{v}}^{\text{final}}$. Image registration calculates solutions and possibly increases $\beta_{\vect{v}}$ internally based on a check of lower admissible bounds for the determinant of the deformation gradient. The deviation of the determinant of the deformation gradient from one indicates local volume changes (expansion for a value larger then one and contraction for a value smaller then one; see \cite{Mang:2016a,Scheufele:2018} for more details).

As a stopping criterion for the Picard iterations, we consider a mixed criterion that is fulfilled either if we reach $\beta_{\vect{v}}^{\text{final}}$ or if the image registration fails due to violation of the user defined lower bound on the determinant of the deformation gradient. We always execute two Picard iterations with the final $\beta_{\vect{v}}^{\text{final}}$.

For both tumor inversion and image registration, we use fixed tolerances for the norm of the reduced gradients as given in~\eqref{e:ma:glob:opt-cond:tumor:inv}, \eqref{e:ma:glob:opt-cond:tumor:inv_diff}, and~\eqref{e:ma:glob:opt-cond:reg:inv}, respectively. Note that we use norms relative to the gradient norm at the beginning of the first Picard iteration for all gradients. We combine these gradient tolerances with a prescribed maximal number of Newton iterations for the reduced space KKT systems (see also~\cite{Scheufele:2018}).

\section{Numerical Methods for Tumor Inversion and Image Registration}
\label{sec_scheuf:components}
In this section, we shortly describe the numerical and algorithmic features of the tumor and the image registration solvers. We give only a brief summary along with references to previous publications since these methods have not been changed from our previous work.

\subsection{Regularization and Nonlinear Solver Components for Tumor Inversion}
\label{sec_scheuf:tumor_reg}

The tumor inversion problem is solved based on an optimize-then-discretize approach. To solve the respective system of optimality conditions, we use different solvers for $\ell_1$ or weighted $\ell_2$-regularization, respectively. For $\ell_1$-regularization, the General Iterative Shrinkage Thresholding (GIST) \cite{Gong2013_GIST} is applied.
In every iteration, it determines a solution $\vect{p}^{k+1}$ close to the result $\vect{p}_{sd}^k$ of a gradient descent step in the direction $\nabla \mathcal{D}_c(\maSimPath, \maPatientPath)$ with step length $\alpha^k$:
\begin{subequations}
	\begin{align}
		\vect{p}^{k+1} &=
		\argmin_{\vect{p}}~ \frac{1}{2}\| \vect{p} - \vect{p}_{sd}^k\|_{L^2(\Omega)}^2 +
		\alpha^k \frac{\beta_{\vect{p}}}{2} \| \vect{p} \|_{L^1(\Omega)} \mbox{,  and  } \\
		\vect{p}_{sd}^k &= \vect{p}^k - \alpha^k\nabla \mathcal{D}_c(\maSimPath,\maPatientPath)
	\end{align}
	To determine the step length $\alpha^k$, we use Armijo line search.
	For $\ell_1$-regularization, we find the analytical solution:
	\begin{align}
		\vect{p}^{k+1} &= \text{sign} (\vect{p}_{sd}^k) \max(0, |\vect{p}_{sd}^k| - \alpha^k \beta_{\vect{p}}).
	\end{align}
\end{subequations}
The stopping criteria are given by
\begin{subequations}
	\begin{align}
		|\mathcal{J}_T(\vect{p}^{k+1}) - \mathcal{J}_T(\vect{p}^k)| &< \tau_J (1 + \mathcal{J}_T(\vect{p}^0))  \\
		\|\vect{p}^{k+1} - \vect{p}^{k}\|_{L^\infty(\Omega)} &< \sqrt{\tau_J} (1 + \|\vect{p}^{k + 1}||_{L^\infty(\Omega)}),
	\end{align}
	where $\mathcal{J}_T$ is the objective function of the tumor inversion problem~\eqref{e:objective_tu}, and $\tau_J > 0$ is a user defined tolerance.
\end{subequations}
We perform parameter continuation in $\beta_{\vect{p}}$ to estimate its value. We start with a large $\beta_{\vect{p}}$, bounded by $\|\nabla \mathcal{D}_c(\maSimPath,\maPatientPath) \|_{L^\infty(\Omega)}$ and perform a binary search in subsequent GIST iterations based on the Hoyer sparsity measure \cite{Hurley2008_HoyerMeasure}, $\mathcal{H}_s (\vect{p}) \in [0, 1]$:
\begin{equation}
	\mathcal{H}_s (\vect{p}) = \frac{n_{\vect{p}} - \nicefrac{\|\vect{p}\|_{L^1(\Omega)}}{\|\vect{p}\|_{L^\infty(\Omega)}}}{n_{\vect{p}} - 1}
\end{equation}
\noindent where $n_{\vect{p}}$ is the dimensionality of $\vect{p}$. Solutions with larger values of $\mathcal{H}_s$ feature higher sparsity, i.e., have less non-zero entries. We use a tolerance of $0.95$ to identify sparse solutions. If the sparsity is large enough (larger than the tolerance), we search for a smaller regularization parameter and, alternatively, increase it if the sparsity is insufficient.
GIST features slow first order convergence, which is, however, acceptable for our purpose as only a rough estimate for $\initcond$ is to be calculated with $\ell_1$-regularization.

In the $\ell_2$-phase, we employ a reduced space method. This means that we assume the state and adjoint equations to be fulfilled exactly; they can be eliminated from the KKT system. For the tumor inversion solver, the reduced gradients are calculated based on the current iterates for $\vect{p}$ and $k_f$ as follows:
\bipa
\item solve the state equations \eqref{e:ma:glob:opt-cond:tumor:state} -- \eqref{e:ma:glob:opt-cond:tumor:state-init},
\item use the result to solve the adjoint equations \eqref{e:ma:glob:opt-cond:tumor:adj} -- \eqref{e:ma:glob:opt-cond:tumor:adj-final},
\item insert the result in \eqref{e:ma:glob:opt-cond:tumor:inv} -- \eqref{e:ma:glob:opt-cond:tumor:inv_diff} to calculate the reduced gradient with respect to $\vect{p}$ and $k_f$.
\eipa

As a solver for the reduced KKT system, we use a quasi-Newton approach (LBFGS, \cite[p.\,135ff]{Nocedal:2006a}) yielding a matrix-free approximation of the matrix-vector product of the gradient with the respective inverse Hessian and, thus, making an inner linear solver obsolete. We use Mor{\'e}-Thuente line-search~\cite{More:1994a} for globalization.\footnote{Note that we do not use line-search in the outer Picard iterations.} The $\ell_2$-phase is terminated when the relative change of the norm of the reduced gradient is below a user defined threshold $\text{opttol}_T>0$. The reference gradient is the gradient obtained for the zero initial guess, $\vect{p}^0$.

\subsection{Nonlinear Solver Components for Image Registration}
\label{sec_scheuf:reg_solve}

We use an optimize-then-discretize approach for the image registration solver~\cite{Mang2018b:CLAIRE}. We employ a reduced space method: The current iterate for the velocity $\vect{v}$ is used to
\bipa
\item solve the state equations \eqref{e:ma:glob:opt-cond:reg:state} -- \eqref{e:ma:glob:opt-cond:reg:state-init} (forward advection),
\item use the result to solve the adjoint equations \eqref{e:ma:glob:opt-cond:reg:adj} -- \eqref{e:ma:glob:opt-cond:reg:adj-final},
\item insert the result in \eqref{e:ma:glob:opt-cond:reg:inv} to calculate the reduced gradient.
\eipa

We use a Newton-Krylov approach with an inexact Gauss-Newton linearization and a matrix-free PCG method as inner linear solver (see~\cite{Mang:2015a,Mang:2016a,Gholami:2017a,Mang:2016c,Mang:2017b} for more details) to compute the search direction.

For the convergence of the optimizer, we use a combination of the relative change of \bipa\item the norm of the gradient in~\eqref{e:ma:glob:opt-cond:reg:inv}, \item the objective in~\eqref{e:objective_reg} and \item the control variable $\vect{v}$, all controlled by a single parameter $\text{opttol}_R>0$, as a stopping criterion. \eipa  More details on the stopping conditions can be found in~\cite{Mang:2015a,Modersitzki:2009a}; see~\cite[305\,ff.]{Gill:1981a} for a discussion. For the inner PCG solver, we perform inexact solves~\cite{Dembo:1983a, Eisenstat:1996a} with a tolerance that is proportional to the norm of the reduced gradient of our problem. This prevents over-solving of the Hessian system if far from the optimum (large gradient norm). See also~\cite[p.~165ff.]{Nocedal:2006a} for details. We use Armijo line-search for globalization.

For both tumor inversion and registration solver, we specify a maximum number ($\text{maxit}_{N,T}$\,/\,$\text{maxit}_{N,R}$) of Newton, a maximum number ($\text{maxit}_{K,T}$\,/\,$\text{maxit}_{K,R}$) of Krylov iterations, and a lower bound of $\num{1E-6}$
for the absolute norm of the gradient as a safeguard against a prohibitively high number of iterations.

\subsection{Discretization in Space and Time for the State and Adjoint Equations}
For all state and adjoint equations in \eqref{e:ma:first_order_opt_global}, we use the following numerical ingredients that have been published in more
detail in ~\cite{Mang:2015a,Mang:2016a,Gholami:2017a,Gholami:2016a,Gholami:2015a,Mang:2016c,Mang:2017b}:
\bipa
  \item spectral elements for spatial discretization combined with a regular mesh;
  \item 3D Fourier transforms to compute spatial derivatives;
  \item an unconditionally stable semi-Lagrangian time-stepping scheme to avoid stability issues and small time-steps for image advection;
	\item an approximation of boundary conditions for the tumor equations at the brain surface based on a penalty approach combined with periodic
	      boundary conditions at the boundary of the overall domain $\bar{\Omega} = [0,2\pi]^3$;
	\item an unconditionally stable, second order Strang-splitting approach~\cite{Strang:1986,Hogea:2008b} with analytical solution of the reaction terms and an implicit Crank-Nicholson method for the diffusion terms in the tumor equations.
\eipa

\section{Numerical Results}
\label{sec_scheuf:results}

The experiments serve as a proof of concept for the newly developed joint registration and biophysical inversion strategy (\textit{Moving Atlas}) and demonstrate that the previously introduced moving patient scheme~\cite{Scheufele:2018} can fail dramatically in terms of both tumor reconstruction and inversion for initial conditions and tumor growth parameters. We quantitatively and qualitatively compare the new \textit{Moving Atlas} strategy to the \textit{Moving Patient} with an emphasis on model inversion and parameter estimation and address the scheme's eligibility to recover true model parameters. This is an important step towards reconstruction of meaningful biophysical characteristics. We consider the following classes of test cases:

\medskip\noindent
\iiparagraph{SYN -- Synthetic Brain} This first academic test case shows that the \ipoint{moving patient scheme} can dramatically fail in terms of biophysical inversion. More precisely, we demonstrate that the moving patient scheme yields largely wrong reconstruction of the tumor, the tumor initial condition, and the diffusion coefficient. We use two simple elliptical anatomies consisting of white matter and gray matter only for atlas and healthy patient brain, respectively, and grow a tumor with known diffusion and reaction parameters in the patient anatomy. The resulting pathologic patient brain serves as target input data for our joint inversion approach. In order to achieve significant results, we create a case with large inter-subject differences between atlas and patient anatomy.

\medskip\noindent
\iiparagraph{STRV -- Synthetic Tumor, Real Velocity} This second class of test cases is based on real brain geometries and serves as a means to
examine the potential of the moving atlas formulation for biophysical inversion in realistic geometries, i.e, to examine the quality of reconstruction for the known ground truth initial conditions, $k_f$ and healthy patient geometry. We use two different real healthy brain geometries and grow a synthetic tumor in one of them, which we use as the patient brain. The second healthy brain serves as a normal brain template (atlas brain). In contrast to real patient tumors, in this case we know the ground truth for tumor initial conditions, diffusion coefficient $k_f$ and healthy patient geometry. Note that we do \textit{not} know the ground truth for the registration velocity $\vect{v}$.

\medskip
In our test cases, we compare three different solution strategies:

\smallskip\noindent\iiparagraph{I. Tumor Standalone (T)}
For the tumor standalone solution, the inversion is carried out assuming a statistical atlas brain geometry as healthy patient brain. No inter-subject registration is applied. The error is a function of the inter-subject variability between atlas and patient brain.

\smallskip\noindent\iiparagraph{II. The Joint Registration and Biophysical Inversion Moving Patient Scheme (MP)}
The \textit{Moving Patient} scheme presented in~\cite{Scheufele:2018} exhibits some inherent shortcomings with respect to the suitability for biophysical inversion and model calibration. The registration may fit the input data to a poor tumor reconstruction. We use the method described in~\cite{Scheufele:2018} enhanced with a sparsity constraint for the tumor inversion solver~\secref{sec_scheuf:tumor_reg}.

\smallskip\noindent\iiparagraph{III. The Joint Registration and Biophysical Inversion Moving Atlas Scheme (MA)}
The \textit{Moving Atlas} scheme presented in this work is designed to remedy some of the shortcomings of the \textit{Moving Patient} scheme.

\medskip
In the following, we shortly describe data and parameters for the experiments in \secref{sec_scheuf:data} as well as the hardware used and the general setup in \secref{sec_scheuf:hardware}. We introduce performance measures used in the evaluation of our numerical results in \secref{sec_scheuf:performance_metrics}. In \secref{sec_scheuf:results_Artificial_Brain}, and \secref{sec_scheuf:results_ATRV}
we show results for the two classes of test cases.

\subsection{Data and Parameters}
\label{sec_scheuf:data}
We use a common data basis for brain geometries as well as some fixed parameter settings for all test cases as described below.

\medskip\noindent
\iparagraph{Brain Imaging Data} For the STRV cases, we use normal brain MR imaging data obtained at the Perelman School of Medicine at the University of Pennsylvania. The three-dimensional imaging data have an image resolution of $256^3$ voxels. We use binary segmentations of MRI scans for white matter (WM), gray matter (GM), and cerebrospinal fluid with ventricles (CSF). In a pre-processing step, these labels are smoothed and rescaled to probability maps, i.e, values between zero and one. To ensure \emph{partition of unity} across all anatomy labels for each $\vect{x}$ in $\Omega$, we introduce a fourth probability map for background based on
\[\forall\vect{x}\in\Omega : c^{(\cdot,1)}(\vect{x}) + \sum_{i=4}^4 m_{i}^{(\cdot,1)}(\vect{x}) \overset{!}{=} 1.\]

\medskip\noindent
\iparagraph{Common Model and Numerical Parameters}
The tumor model and solver parameters are summarized in~\tabref{tab:tc-parameters}. The solver parameters are based on experiments. We briefly discuss some of the more involved choices:
For the parametrization of the tumor initial condition, we select a regular grid of $n_{\vect{p}}$ Gaussian basis functions with a variable standard variation $\sigma \in \{\nf{2\pi}{30}, \nf{2\pi}{64}\}$ and a grid spacing of $\delta=1.5\sigma$ around the center of mass of the tumor. The number $n_{\vect{p}}$ is chosen a priori on a case-by-case basis, such that the pathological domain is covered sufficiently.

For all experiments, we consider tumor growth only in white matter ($k_1 =1; \rho_1 = 1$), and, consequently, set the characteristic diffusivity and net cell proliferation in gray matter and CSF to zero, i.e., $k_2=k_3=0$, and $\rho_2=\rho_3=0$. To grow the synthetic tumor, we choose $\rho_f^\star = 15$ as cell proliferation rate (reaction coefficient), and $k_f^\star=\num{1E-1}$ as rate of cell migration into surrounding tissue (diffusion coefficient), respectively. For the tumor evolution simulation, we choose a time step of $\Delta t = 0.01$ and various time horizons $T \in \{0.16, 0.32, 0.44\}$.
The regularization parameter $\beta_{\vect{p}}$ for the $\ell_2$-regularized inverse tumor problem has been determined experimentally from an L-curve study similar to~\cite{Gholami:2016a}, using $n_{\vect{p}}=125$ Gaussian basis functions and an image resolution of $n=128^3$. We fix the relation $\nf{\sigma}{\delta}=1.5$ for the spacing of the Gaussian basis functions\footnote{This leads to an invariant condition number of $\Phi^T\Phi$ for the corresponding interpolation problem for varying standard variations $\sigma$, i.e., width of the Gaussians.}.
The lower bound $\beta_{\vect{v}}^{final}$ on the regularization parameter $\beta_{\vect{v}}$ has been determined based on extensive numerical experiments for different synthetic and real brain data sets~\cite{Mang:2016a,Mang:2017b}. By experimental analysis we found that reducing the regularization weight by a factor of $10$ in every Picard-iteration is sufficient\footnote{Using a smaller reduction or more Picard-iterations per fixed regularization weight did not improve the overall result.}.
For the optimizer, we use tolerances of $\mbox{opttol}_R = \mbox{opttol}_T = \num{1E-3}$ for registration and tumor inversion, respectively\footnote{Note that, in combination with the applied stopping conditions, this results in a required relative gradient of $\num{1E-3}$ for the tumor inversion, but a gradient reduction of only about two orders of magnitude for the registration.}. Further reduction of the gradient did not improve the final reconstruction quality, since the overall error is bound by an $\mathcal{O}(1E-3)$ error introduced by the solution of the hyperbolic advection problem (cf.~\cite{Scheufele:2018}). To prevent prohibitively large runtimes, we define upper bounds on the number of Newton (and Krylov) iterations.

\begin{table}[htb]
\caption{\label{tab:tc-parameters} Summary of common parameters used in all test cases. We report values of the following parameters: $n$ denotes the image resolution with $n_i$, $i=1,2,3$; $n_{\vect{p}}$ is the number of Gaussian for the parametrization of the tumor initial condition, $\sigma$ is the standard deviation of the associated Gaussian basis functions (and varies throughout test cases), $\delta$ denotes the spacing in between centers of adjacent Gaussians; $k_1$, $k_2$ and $k_3$ are the characteristic diffusion parameters for WM, GM, and CSF, $k_f$ the overall scaling parameter for the isotropic part of the diffusion coefficient for net migration of cancerous cells into surrounding tissue; $\rho_1$, $\rho_2$, and $\rho_3$ are the characteristic reaction factors for WM, GM, and CSF, $\rho_f$ is the overall reaction scaling factor;
$\text{opttol}_R$, $\text{opttol}_T$ are the convergence tolerances for registration and tumor inversion; $\text{maxit}_i = (\text{maxit}_{i,N}, \text{maxit}_{i,K})$ denotes the maximum number of Newton iterations and Krylov iterations (for the KKT system) for the tumor inversion ($i=T$) and registration ($i=R$), respectively; $\beta_{\vect{p}}$ is the regularization parameter for the tumor inversion; $\beta_{\vect{v}}^{init}$ and $\beta_{\vect{v}}^{final}$ are the initial and final values for the $\beta$-continuation scheme, applied in image registration; $\varepsilon_{\igrad}$ is the bound on the variation of the deformation gradient $\mbox{det}(\igrad \vect{y})$ used in the continuation scheme.
 }
\small\centering
\begin{tabular}{lll}
\toprule
\textbf{Description} & \textbf{Parameter(s)} & \textbf{Value(s)} \\
\midrule
\textit{image resolution}                                              & $n = n_1 \cdot n_2 \cdot n_3$        & $n_i = 128$                         \\
\textit{tumor initial condition parametrization}                       & $n_{\vect{p}}$, $\sigma$, $\delta$   & $\{125,343\}$,  $\{\frac{2\pi}{30}, \frac{2\pi}{64}\}$, $1.5$                  \\
\textit{tumor growth characteristic cell diffusivity}                  & $\bar{k} = k_f^\star(k_1,k_2,k_3)^T$ & $0.1\cdot(1,0,0)^T$                 \\
\textit{tumor growth characteristic net cell proliferation}            & $\bar{\rho} = \rho_f^\star(\rho_1,\rho_2,\rho_3)^T$ & $15\cdot(1,0,0)^T$   \\
\textit{optimizer tolerances}                                          & $\text{opttol}_R$, $\text{opttol}_T$ & $\num{1E-3}$, $\num{1E-3}$          \\
\textit{maximum Newton\,/\,Krylov iterations, tumor inversion}         & $\mbox{maxit}_T = (\text{maxit}_{T,N}, \text{maxit}_{T,K})$ & $(50,-)$   \\
\textit{maximum Newton\,/\,Krylov iterations, registration}            & $\mbox{maxit}_R = (\text{maxit}_{R,N}, \text{maxit}_{R,K})$ & $(50,80)$    \\
\textit{tumor inversion regularization parameter}                      & $\beta_{\vect{p}}$                   & $\num{1E-4}$                        \\
\textit{registration regularization parameter (continuation)}          & $\beta_{\vect{v}}^{init}$, $\beta_{\vect{v}}^{final}$ & $1$, $\num{1E-04}$ \\
\textit{bound on local volume change (variation $\mbox{det}(\igrad \vect{y})$)}  & $\varepsilon_{\igrad}$     & $\num{1E-2}$                        \\
\bottomrule
\end{tabular}
\end{table}

\medskip\noindent
\subsection{Hardware and Setup}
\label{sec_scheuf:hardware}
All numerical experiments were executed on the supercomputer \hazelhen\; at the High Performance Computing Center HLRS in Stuttgart (\url{www.hlrs.de}), a Cray XC40 system with a peak performance of $7.42$ Petaflops comprising $7,712$ nodes with Xeon E5-2680 v3 processors and $24$ cores on two sockets and 128 GB memory per node. Our \sibia~framework is written in \texttt{C++}  and uses MPI for parallelism. It is compiled using the Intel 17 compiler. We use PETSc's implementations for linear algebra operations and PETSc's TAO package for the nonlinear optimization~\cite{petsc-user-ref,Munson:2015a}, AccFFT for Fourier transforms~\cite{Gholami:2016a,accfft-home-page}, and PnetCDF for I/O~\cite{pnetcdf_web}. We use $3$ nodes with $64$ MPI tasks for (down-sampled) data sizes of size $n_i = 128$, $i=1,2,3$ for all runs reported in this study.
%

\subsection{Performance Metrics}
\label{sec_scheuf:performance_metrics}
The task of our numerical experiments is to assess
\bipa
\item the convergence towards solutions with low mismatch between the prediction of our model in~\eqref{e:tumor:fwd}--\eqref{eq:reg:divfree} and the observed patient input data, both in the brain geometry $\vect{m}$ and the tumor $c$ (for all test cases),
\item the reconstruction quality for the (in practice inaccessible) healthy patient brain\footnote{For our semi-synthetic test case setting, where the patient target data is generated from synthetic tumor progression simulation in the healthy brain, the healthy patient brain is known and the approximation quality can be measured exactly.} as a direct output of the moving atlas formulation,
\item the quality of inversion for the biophysical parameters, i.e., tumor growth initial conditions and diffusion coefficient, in terms of correctness of the parameters compared to the known ground truth.
\eipa

We report all performance measures in the patient space, i.e., with respect to the patient anatomy and not the atlas anatomy, since the patient space is the relevant space from an applications point of view. We use the following metrics: We measure the relative mismatch/residual between patient anatomy and reconstructed anatomy, between the reconstructed healthy patient anatomy and the ground truth, and between patient tumor and simulated tumor in the patient domain (compare~\figref{tab:notation}):
\begin{equation*}
  \renewcommand{\arraystretch}{1.2}
\begin{tabular}{p{5cm}p{5cm}p{5cm}}
$\elltwoB \defeq \frac{\| \maSimBrain - \maPatientBrain \|_{L^2(\Omega)^3}}{\| \maHealthyAtlas - \maPatientBrain \|_{L^2(\Omega)^3}},$
&
$\elltwoHPB \defeq \frac{\| \maHealthyPatient - \vect{m}_P^\star(1) \|_{L^2(\Omega)^3}}{\| \maHealthyAtlas - \vect{m}_P^\star(1) \|_{L^2(\Omega)^3}},$
&
$\elltwoT \defeq \frac{\| \maSimPath - \maPatientPath \|_{L^2(\Omega)^3}}{\| \maPatientPath \|_{L^2(\Omega)^3}}$
\end{tabular}
\end{equation*}
where $\vect{m}_P^\star(1)$ denotes the ground truth healthy patient brain.
Based on the cardinality $|\,\cdot\,|$ of a set and a selection function $H$ with threshold $0.5$, i.e.,
\[
H(\vect{u}) \defeq \{ u_i \geq 0.5 \},
\]
we calculate Dice coefficients for the individual label maps  associated with the anatomy labels for $l
\in \{WM, GM, CSF \}$ and their average across labels for the patient and the atlas anatomy
\begin{equation*}
  \renewcommand{\arraystretch}{1.2}
\begin{tabular}{p{7cm}p{7cm}}
$\text{DICE}_{l,B} \defeq 2\frac{|H(m_{P,l}(1)) \cap H(m_{D,l})|} {|H(m_{A,l})| + |H(m_{D,l})|},$
&
$\diceB = \sum_{l=1}^3\text{DICE}_{l,B} / 3;$
\end{tabular}
\end{equation*}
\noindent
Analoguously, we report values for the Dice coefficient computed for the healthy patient geometry, denoted by
$\text{DICE}_{B_0}$, and for the anatomy label of the tumor, denoted by $\text{DICE}_T$. To monitor convergence of the Picard iterations, we report the relative value of an \textit{approximation} to the gradient for the coupled problem (see~\secref{sec_scheuf:ma_formulation}) for the final iteration $k$:
\[
\relG
\defeq \|\vect{g}^j\|_{L^2(\Omega)} / \|\vect{g}^0\|_{L^2(\Omega)}.
\]
where $\vect{g}^j$ is an approximation to the gradient of the coupled optimization problem~\eqref{eq:ma:global_opt} (given by
\eqref{e:ma:glob:opt-cond:tumor:inv} -- \eqref{e:ma:glob:opt-cond:reg:inv}) after the $i$th Picard iteration and $\vect{g}^0$ the corresponding approximate gradient for the initial guess. The gradient is only approximate, since we neglect the term $\partial_{\maHealthyPatient} \Phi(\vect{p}) \alpha(0)$ in the final condition~\eqref{e:ma:glob:opt-cond:reg:adj-final} of the adjoint equation for the registration problem. This term is guaranteed to be small if the norm of the remaining gradient is small as the latter requires $\alpha(0)$ and the change in the estimated patient
geometry to be small.
Finally, we calculate the relative $\ell_2$-error for the initial condition:
\[
\elltwoINIT \defeq \|\Phi \vect{p}^\star - c(0) \|_{L^2(\Omega)} /\| \Phi \vect{p}^\star \|_{L^2(\Omega)},
\]
where $c(0) = \Phi \vect{p}$ and $ \Phi \vect{p}^{\star}$ denotes the parametrization of the ground truth tumor initial condition.

\subsection{Results for SYN -- Synthetic Brain}
\label{sec_scheuf:results_Artificial_Brain}
This first, purely synthetic experiment uses two fairly different, very simplistic anatomies for the atlas and patient brain and an artificially grown tumor in one of them (called the patient brain).

\medskip\noindent
\iparagraph{Purpose} This case is designed to cause the \textit{Moving Patient} scheme to fail due to large differences in the brain anatomy between patient and atlas in the tumor region. We demonstrate that the \textit{Moving Atlas} scheme retains good reconstruction quality for the tumor anatomy label, the tumor initial condition, and the characteristic diffusivity of tumor cells, while the \textit{Moving Patient} scheme falls short as it inverts tumor growth in the 'wrong' (atlas) geometry.

\begin{figure}[htb]\centering
  \begin{minipage}[t]{0.495\textwidth}\centering
    \footnotesize\emph{Reconstruction using the Moving Atlas Scheme} \\
  \includegraphics[width=\textwidth, trim=0 0 0 0, clip]{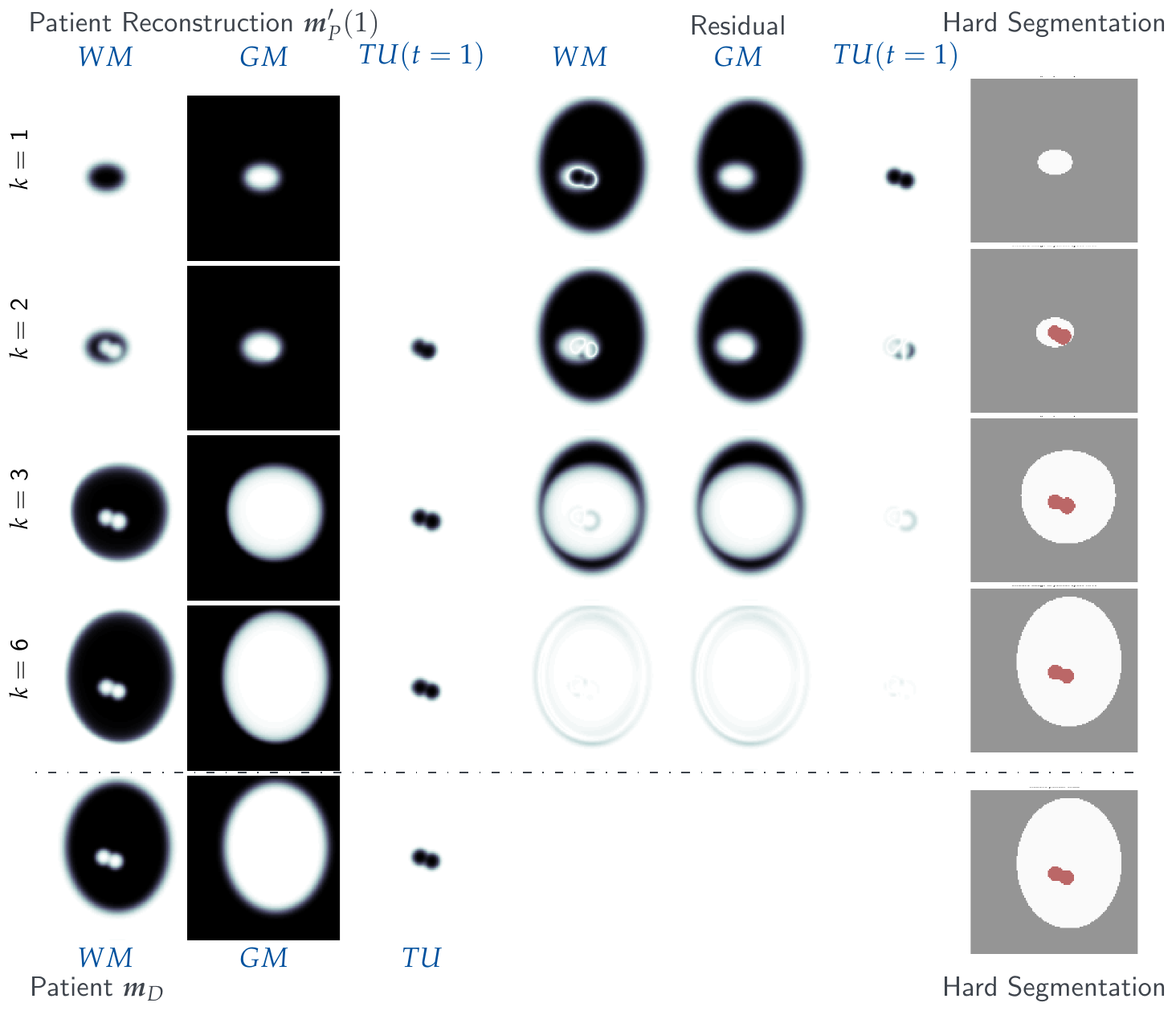}
  \end{minipage}
  \begin{minipage}[t]{0.495\textwidth}\centering
    \footnotesize\emph{Reconstruction using the Moving Patient Scheme} \\
  \includegraphics[width=\textwidth, trim=0 0 0 0, clip]{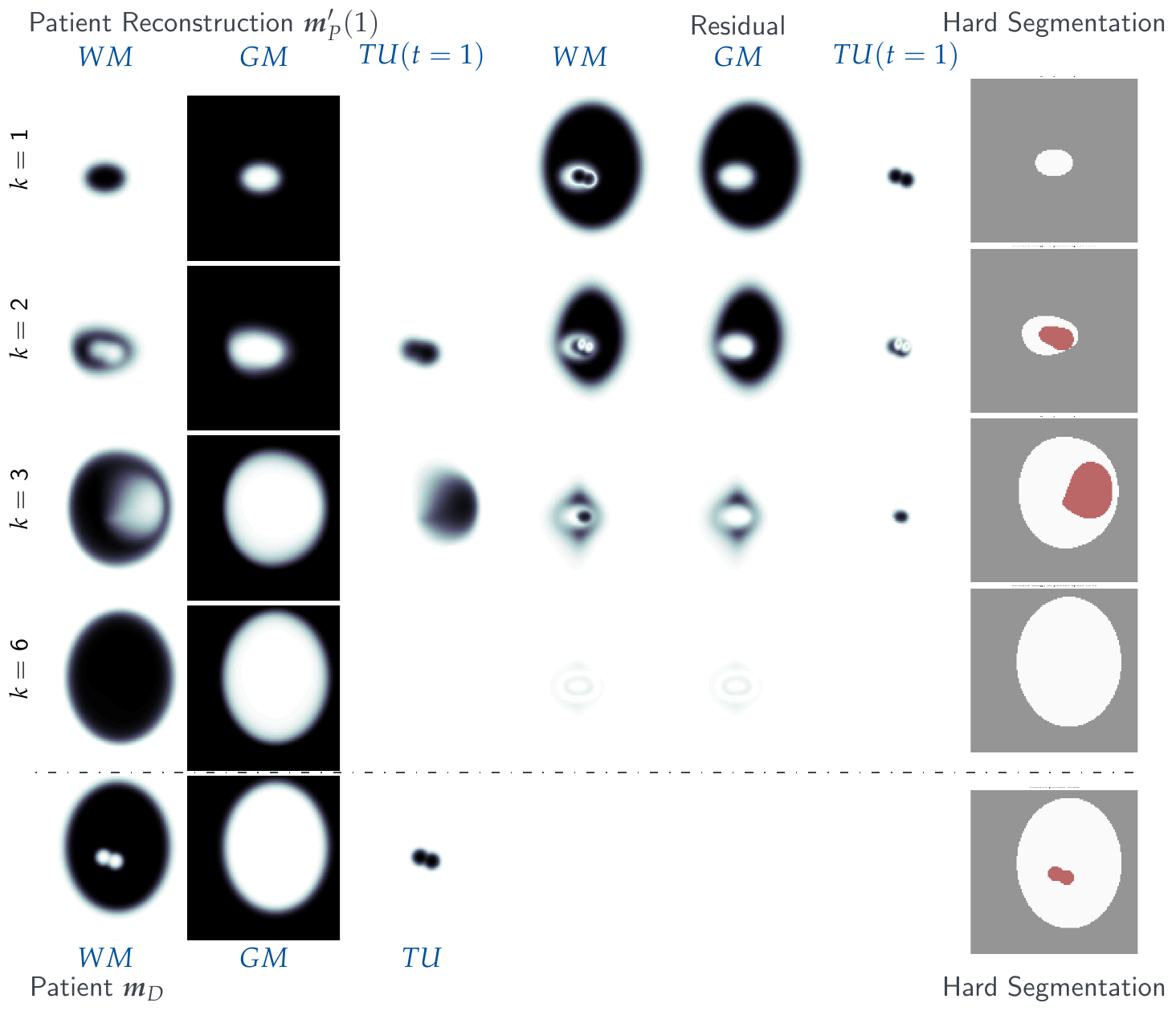}
  \end{minipage}
  \\
  {
  \vspace{1pt}
  \centering
  \footnotesize\emph{Quantitative Results} \\
  \begin{small}
  \setlength\tabcolsep{4pt}
  \begin{tabular}{r|LlLlLlLlL}
  \toprule
  & inv-$k_f$ &  $\errDiffInversion$ & $\elltwoB$  & $\diceB$ & $\elltwoHPB$  & $\diceHPB$ & $\elltwoT$  & $\diceT$  & $\elltwoINIT$ \\
  \midrule
  \textbf{\textit{MA}}   & \num{7.681e-3} & \num{2.32e-1} & \num{5.038000e-02}  & \num{9.976500e-01} & \num{3.253000e-02}  & \num{9.977000e-01} & \num{3.321000e-02}  & \num{9.891000e-01}  & \num{1.836000e-01} \\
  \textbf{\textit{MP}}   &  \num{0.0}  & \num{1.00e+00} & \num{6.495000e-02}  & \num{9.953000e-01} & \num{3.482000e-02}  & \num{9.962000e-01} & \num{9.880000e-01}  & \num{0.000000e+00}   & \num{1.00e-00} \\
  \bottomrule
  \end{tabular}
\end{small}
  }
  \caption{Results for \ipoint{SYN -- Synthetic Brain} (ground truth proliferation rate $\rho_f=15$ and diffusivity $k_f=\num{1E-02}$ in white matter,
	tumor growth disabled in gray matter). We compare the new \ipoint{Moving Atlas (MA)} Picard scheme (left tableau), and the \ipoint{Moving Patient (MP)} scheme (right tableau). The figure shows anatomy and tumor labels (from left to right: WM, GM, CSF, and TU), the residual differences (if available), and a hard segmentation based on the given probabilities for the individual tissue classes. The synthetic healthy atlas and the patient are composed of an elliptic inclusion of white matter in a gray matter rectangular brain (see text for details; axial-slice $64$ of a 3D volume). We show the initial configuration for the problem (top row; iteration $k=1$), two intermediate results (second row; $k=2$, and third row; $k=3$), the final configuration after joint registration and tumor inversion (fourth row; iteration $k=6$) and the target patient data (reference image; bottom row). All results are presented in patient space.
  In the table, we report the reconstructed diffusivity inv-$k_f$ and its relative error $\errDiffInversion$ with respect to the ground truth, the average mismatch for the anatomy labels of the tumor bearing brain $\elltwoB$, for the healthy brain tissue labels $\elltwoHPB$ and for the tumor $\elltwoT$, the mean Dice coefficients for brain tissue $\diceB$, for the healthy patient brain tissue $\diceHPB$ and for the tumor $\diceT$.
  Furthermore, we measure the $\ell_2$-error $\elltwoINIT$ of the reconstructed tumor initial condition with respect to the ground truth.
  \label{fig:SYN-tc-ma-versus-mp}
  }
\end{figure}

\medskip\noindent
\iparagraph{Setup}
For both, atlas and patient, we generate a purely synthetic brain anatomy, composed of an elliptical white matter inclusion surrounded by gray matter. The white matter inclusions for atlas and patient differ greatly in size. The target data $\maPatientPath$ are generated synthetically from a tumor-progression simulation (with $\vect{p} = \vect{p}^\star$) using our reaction-diffusion model with $\rho_f^\star=15$ and $k_f^\star=\num{1E-2}$ from an initial condition with two nearby Gaussians enabled
in the patient geometry. The healthy atlas geometry is depicted in the first row, the tumor bearing patient geometry in the last row of~\figref{fig:SYN-tc-ma-versus-mp}. We use the same set of $n_{\vect{p}}=343$ Gaussians with standard deviation $\sigma=\nf{2\pi}{30}$ for the target data generation and inversion. For simplicity, tumor growth is enabled only in white matter. We invert for the registration velocity $\vect{v}$, the parameters $\vect{p}$ for the tumor initial condition, and for the characteristic diffusivity $k_f$ in white matter.

\medskip\noindent
\iparagraph{Observation}
In this case, the inclusions of white matter in atlas and patient vary significantly in size (cf.\;\figref{fig:SYN-tc-ma-versus-mp}). Thus, registering the atlas to the patient requires strong local expansion of volume. Looking at~\figref{fig:SYN-tc-ma-versus-mp} (right tableau), in the first iteration, to prevent the registration from directly matching the target tumor unaided by the tumor solver, we choose a higher regularization parameter $\beta_{\vect{v}}$ for the registration problem. Therefore, warping the target tumor to the atlas space basically corresponds to a copy operation. (before $k=2$). This warped-to-atlas tumor is used for tumor inversion in the atlas space. The subsequent registration step (with lower regularization) computes a velocity that registers the patient (small tumor in large white matter blob) to the atlas (small tumor in small white matter blob). This results in a compression of the patient white matter, and consequently yields a very small target tumor for the inversion in the atlas ($k=3$).\footnote{Note, the displayed results are in the patient space, i.e., the inverse of the deformation is applied to display atlas labels in patient space, and the very small reconstructed tumor in the atlas is expanded when warped to the patient space.} In subsequent iterations (with further reduced registration regularization), the warped-to-atlas target tumor is compressed further, ultimately causing the \textit{Moving Patient} scheme to fail and returning a relative error of almost $100\%$ for the reconstruction of the tumor label, the reconstruction of the tumor initial condition, and the reconstruction of the characteristic diffusivity $k_f$ in white matter cf.\;\figref{fig:SYN-tc-ma-versus-mp}). The new \textit{Moving Atlas} scheme, however, maintains a good reconstruction quality and results in only $3.3\%$ relative errors for the tumor label, $18.4\%$ relative errors for the reconstruction of the tumor initial condition and $23.2\%$ relative error in the reconstruction of the characteristic diffusivity rate $k_f$ in white matter.

\subsection{Results for STRV -- Synthetic Tumor \& Real Velocity}
\label{sec_scheuf:results_ATRV}

With the results for the STRV test cases, we show the general applicability of the new registration and tumor inversion coupling scheme as well as its improved properties in terms of correct reconstruction of initial tumor and healthy brain geometry. We evaluate the \textit{Moving Atlas} scheme for a real brain geometry with synthetically generated target tumor such that the true tumor model parameters are known and we have access to the true healthy patient brain.

\medskip\noindent
\iparagraph{Purpose} With the STRV test case experiments, we pursue three main purposes:
\bipa
\item We give experimental evidence for the convergence of the new \textit{Moving Atlas} scheme for several settings;
\item we compare our coupled image registration and tumor inversion schemes \textit{Moving Patient} and \textit{Moving Atlas} to standalone tumor inversion, replacing the missing second snapshot (healthy patient) by a statistical atlas without involving image registration;
\item we show the superiority of the new \textit{Moving Atlas} scheme over the previously presented \textit{Moving Patient} scheme in terms of reconstruction quality of the
(initially unknown) healthy patient brain anatomy and meaningful reconstruction of biophysical tumor-growth characteristics, in particular shape and sparsity of the ground truth tumor initial condition and prediction of the grown tumor.
\eipa

\medskip\noindent
\iparagraph{Setup}
For both atlas and patient, we use real brain MR imaging data from two (healthy) individuals. The target tumor $\maPatientPath$ is generated synthetically from a tumor-progression simulation (with $\vect{p} = \vect{p}^\star$) using our reaction-diffusion model. Accordingly, the ground truth velocity field is unknown (or may not even exist due to a possibly wrong model), but the true tumor model parameters are known.
We consider two different tumor locations (STRV-C1) and (STRV-C2) and varying time horizons for the tumor-progression simulation. An overview is given in~\tabref{c8:biophysics:atrv:testcase-setup}.
For (STRV-C1) we also employ different settings of the parametrization of the tumor initial condition: For parameter setting $\#1$, we use a set of $n_{\vect{p}}=125$ Gaussian basis functions with standard deviation $\sigma=\nf{2\pi}{30}$, whereas, for settings $\#2$ and $\#3$, we use smaller (but more; $n_{\vect{p}}=343$) Gaussians with $\sigma=\nf{2\pi}{64}$ to generate sparser ground truth initial conditions. The same set of Gaussian basis functions is used for target data generation and inversion.
The target data are generated from a forward simulation with $\rho_f^\star=15$ and $k_f^\star=\num{1E-1}$ and an initial condition with two nearby Gaussians enabled. For simplicity, tumor growth is enabled only in white matter. We invert for the tumor initial condition and the characteristic diffusivity $k_f$ in white matter.
%
\begin{table}[htb]
  \small\centering
  \caption{STRV test case variations. Parameter choices for the \ipoint{synthetic tumor with real velocity (STRV)} test case;  ground truth: ($\rho^{\star}_f=15$, $\rho^{\star}_1=1$, $\rho^{\star}_1=0$, $k^{\star}_f=\num{1.0E-1}$, $k^{\star}_1=1$, $k^{\star}_1=0$, $\vect{p}=\vect{p}^\star$). For the initial condition parametrization, we use a regular grid of $n_{\vect{p}} = 125$ (large $\sigma$), and $n_{\vect{p}} = 343$ (small $\sigma$) Gaussian basis functions, respectively with standard deviation $\sigma$ as outlined below, and a spacing of $\delta=1.5\,\sigma$. The grid is centered around the positions $\vect{x}_{c_1}=\nf{2\pi}{128}(39, 63, 64)$ (STRV-C1) and $\vect{x}_{c_2}=\nf{2\pi}{128}(61, 89, 64)$ (STRV-C2), respectively (cf.\;\figref{fig:c8:biophysics:atrv-cm3-ground-truth-T-032-and-T-16-k-01-sigma-pi-32-SIM-TC-large-TOGETHER} and~\figref{fig:c8:biophysics:atrv-cm5-ground-truth-T-032-and-T-16-k-01-sigma-pi-32-SIM-TC-large-TOGETHER} for the location of tumor seed).
  }
  \label{c8:biophysics:atrv:testcase-setup}
\begin{tabular}{l|ccc}
  \toprule
  \textbf{test case}                         & \textbf{setting \#1}             & \textbf{setting \#2}             & \textbf{setting \#3} \\
  \midrule
  \textbf{(STRV-C1)}  & $\sigma=\nf{2\pi}{30}$, $T=0.16$ & $\sigma=\nf{2\pi}{64}$, $T=0.32$ & $\sigma=\nf{2\pi}{64}$, $T=0.44$ \\
  \textbf{(STRV-C2)}  & $\sigma=\nf{2\pi}{64}$, $T=0.16$ & $\sigma=\nf{2\pi}{64}$, $T=0.32$    \\
  \bottomrule
\end{tabular}
\end{table}

\begin{figure}\centering
  \emph{Qualitative Results for the Moving Atlas Joint Inversion Scheme} \\
  \makebox[\textwidth]{\centering
  \includegraphics[width=\textwidth, trim=0 0 0 0, clip]{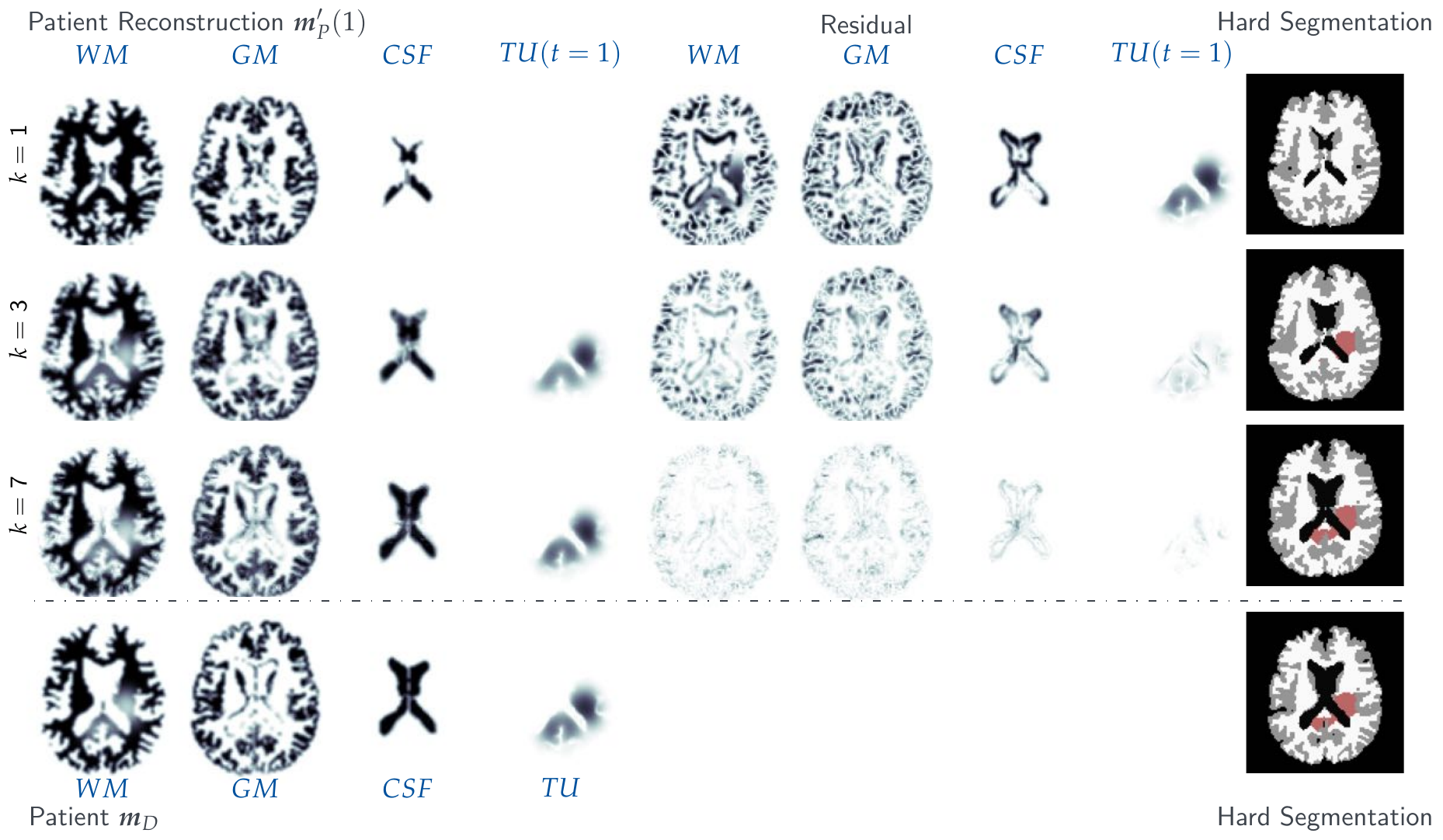}}
  \\
  {
  \centering
  \emph{Quantitative Results for the Moving Atlas Joint Inversion Scheme}
  \begin{small}
  \setlength\tabcolsep{4pt}
  \begin{tabular}{rr|LlLlLllLl}
  \toprule
  It      & $\beta_{\vect{v}}$ & $\elltwoB$  & $\diceB$ & $\elltwoHPB$  & $\diceHPB$ & $\elltwoT$  & $\diceT$    & $\relG$  & $\elltwoINIT$   & $\rtIT$\,[s] \\
  \midrule
  $  1$ &  \num{1e+00}  & \num{1.000000e+00}  & \num{5.450333e-01} & \num{6.773000e-01}  & \num{5.505667e-01} & \num{1.000000e+00}  & \num{0.000000e+00}  & \num{1.000000e+00}  & \num{1.000000e+00}  & \num{5.641000e+02}   \\
  $  2$ &  \num{1e-01}  & \num{8.791000e-01}  & \num{5.798333e-01} & \num{6.067000e-01}  & \num{5.928000e-01} & \num{2.899000e-01}  & \num{7.921000e-01}  & \num{2.037000e-02}  & \num{4.252000e-01}  & \num{9.451000e+02}   \\
  $  3$ &  \num{1e-02}  & \num{6.538000e-01}  & \num{7.286333e-01} & \num{4.508000e-01}  & \num{7.379000e-01} & \num{2.037000e-01}  & \num{8.236000e-01}  & \num{1.450000e-02}  & \num{3.511000e-01}  & \num{8.441000e+02}   \\
  $  4$ &  \num{1e-03}  & \num{4.062000e-01}  & \num{8.662667e-01} & \num{2.801000e-01}  & \num{8.721000e-01} & \num{1.157000e-01}  & \num{8.804000e-01}  & \num{1.362000e-02}  & \num{2.081000e-01}  & \num{1.627000e+03}   \\
  $  5$ &  \num{1e-04}  & \num{2.620000e-01}  & \num{9.324667e-01} & \num{1.813000e-01}  & \num{9.365000e-01} & \num{7.573000e-02}  & \num{9.483000e-01}  & \num{7.010000e-03}  & \num{9.548000e-02}  & \num{8.462000e+02}   \\
  $  6$ &  \num{1e-04}  & \num{2.364000e-01}  & \num{9.416333e-01} & \num{1.638000e-01}  & \num{9.455000e-01} & \num{6.740000e-02}  & \num{9.511000e-01}  & \num{4.905000e-03}  & \num{5.754000e-02}  & \num{8.577000e+02}   \\
  \midrule
  $  7$ &  \num{1e-04}  & \num{2.333000e-01}  & \num{9.431667e-01} & \num{1.616000e-01}  & \num{9.474333e-01} & \num{6.491000e-02}  & \num{9.498000e-01}  & \num{4.153000e-03}  & \num{7.134000e-02}  & \num{5.688502e+03}   \\
  \bottomrule
  \end{tabular}
\end{small}
  }
  \caption{Misfit reduction of the new \ipoint{Moving Atlas (MA)} Picard iteration scheme. We show qualitative and quantitative results for the (STRV-C1) test case with parameter setting $\#1$ from~\tabref{c8:biophysics:atrv:testcase-setup} over the course of the \ipoint{Moving Atlas} Picard-iteration-type solution scheme with a sparsity constraint on the initial condition for the tumor inversion solver. We use the ground truth values $\rho_f = \rho_f^\star=15$ and $k_f = k_f^\star=\num{1e-1}$ in white matter for the inversion.
  The figure shows anatomy labels of the healthy atlas brain and the patient brain with an (artificially) grown tumor (see text for details; axial-slice $64$ of a 3D volume). We show the initial configuration for the problem (top row; iteration $k=1$), an intermediate result (second row; $k=3$), the final configuration after joint registration and tumor inversion (third row; iteration $k = 7$), and the target patient data (reference image; bottom row). Each row contains (from left to right) the anatomy labels for WM, GM, CSF, and TU, the residual differences (if available) between the anatomy labels, and a hard segmentation based on the given probabilities for the individual tissue classes.
  In the table, we report the average mismatch for the anatomy labels for the pathologic brain tissue labels $\elltwoB$, the healthy brain tissue labels $\elltwoHPB$ and the tumor $\elltwoT$, the mean Dice coefficient for brain tissue $\diceB$, healthy patient brain tissue $\diceHPB$ and tumor $\diceT$ over the Picard iterations. Furthermore, we measure the $\ell_2$-error $\elltwoINIT$ of the reconstructed tumor initial condition to the ground truth, and the relative norm  $\relG$  of the approximated reduced gradient of the coupled formulation. Timings per iteration are given in seconds for parallel execution using 64 MPI tasks on three nodes of~\hazelhen. The last row shows the final values for mismatch and Dice as well as the time-to-solution (accumulated runtime) in seconds.
  \label{fig:c8:biophysics:atrv-cm3-ground-truth-T-016-k-01-sigma-pi-32_moving-atlas-run-over-iterations}
  }
\end{figure}

\medskip\noindent
\iparagraph{Misfit and Gradient Reduction of the Moving Atlas Scheme} To experimentally assess convergence of the \textit{Moving Atlas} scheme over the
Picard iterations, we report the $\ell_2$-mismatch and Dice overlay coefficients for the reconstruction of the (pathologic) brain anatomy ($\elltwoB$ and $\diceB$), the healthy patient anatomy ($\elltwoHPB$ and $\diceHPB$), and the reconstructed (grown) tumor ($\elltwoT$ and $\diceT$) in~\figref{fig:c8:biophysics:atrv-cm3-ground-truth-T-016-k-01-sigma-pi-32_moving-atlas-run-over-iterations} for the (STRV-C1) test case. Furthermore, we monitor the $\ell_2$-error $\elltwoINIT$ for the reconstructed tumor initial condition with respect to the ground truth and the relative norm $\relG$ of the (approximated) reduced gradient of the coupled formulation~\eqref{eq:ma:global_opt}.

\medskip\noindent
\iiparagraph{Observations} \figref{fig:c8:biophysics:atrv-cm3-ground-truth-T-016-k-01-sigma-pi-32_moving-atlas-run-over-iterations} shows a monotonic reduction of data-misfit values and the corresponding gain in tissue overlay Dice scores for the brain anatomy, the healthy patient anatomy, the grown tumor, and the initial tumor condition. Although we do not have a proof for the convergence of the \textit{Moving Atlas} Picard-iteration-type solution strategy (outlined in~\secref{sec_scheuf:picard}), we monitor the relative norm $\relG$ of the (approximated) reduced gradient for the \textit{Moving Atlas} formulation~\eqref{eq:ma:global_opt}, indicating convergence to an optimum. The norm of the approximated gradient continually decreases throughout our Picard-iteration-type solution strategy. Our solution scheme, and hence the gradient, neglect the term $\partial_{\maHealthyPatient} \Phi(\vect{p}) \alpha(0)$ in~\eqref{e:ma:glob:opt-cond:reg:adj-final}. This term, however, becomes very small as the misfit and, thus, $\alpha(0)$ decrease.

\resetrunid
{
\setlength\tabcolsep{3pt}
\begin{table}
  \caption{
Comparison of the \ipoint{tumor inversion standalone (T)} solver only (w/o inter-subject registration), the \ipoint{Moving Patient (MP)} Picard iteration scheme, and the new \ipoint{Moving Atlas (MA)} Picard iteration scheme. We show quantitative results for the (STRV-C1) and (STRV-C2) test cases with centers $\vect{x}_{c_1}$ and $\vect{x}_{c_2}$, and parameter setting $\#1$ from~\tabref{c8:biophysics:atrv:testcase-setup}, respectively. We invert for the registration velocity $\vect{v}$, the parameters $\vect{p}$ of the tumor initial condition, and for the characteristic diffusivity $k_f$ in white matter. We set the proliferation rate to the true value $\rho_f=15$. We report the reconstructed diffusivity inv-$k_f$ and its relative error $\errDiffInversion$ with respect to the ground truth, the average mismatch for the anatomy labels for the pathologic brain tissue labels $\elltwoB$, the healthy brain tissue labels $\elltwoHPB$ and the tumor $\elltwoT$, the mean Dice coefficients for brain tissue $\diceB$, for the healthy patient brain tissue $\diceHPB$ and for the tumor $\diceT$. We also measure the $\ell_2$-error $\elltwoINIT$ of the reconstructed tumor initial condition to the ground truth. Timings are given for parallel execution using 64 MPI tasks on three nodes of \hazelhen. 
  \label{tab:ATRV-DIF-CM-3::T-016-dt-001-k-01-sigma-pi-15::MA-vs-MP-L1wL2-vs-L2-simple-obj}
}
\centering
\begin{scriptsize}
\makebox[\textwidth]{\centering
\begin{tabular}{llll|lLLlLlLlLlll}
\toprule
ID & center & setting & solver & inv-$k_f$ & $\errDiffInversion$ & $\elltwoB$  & $\diceB$ & $\elltwoHPB$  & $\diceHPB$ & $\elltwoT$  & $\diceT$  & $\elltwoINIT$  & $\rtIT$\,[s] & $\rtTUM$\,[s] & $\rtREG$\,[s]  \\
\midrule
$\runid$ & \multirow{3}{*}{{\scriptsize \centering\textbf{\textit{C1}}}}
& \multirow{3}{*}{{\scriptsize \centering\textbf{\textit{\#1}}}}
& \textbf{\textit{T}} & \num{1.149e-01} & \num{1.5e-01} & N/A & N/A & N/A & N/A & \num{2.899000e-01} & \num{7.921000e-01} & \num{4.252000e-01} & N/A & -- & N/A  \\
$\runid$ & & & \textbf{\textit{MP}} & \num{9.222e-02} & \num{7.8e-02} & \num{3.715000e-01}  & \num{8.724000e-01} & \num{2.555000e-01}  & \num{8.793000e-01} & \num{1.442000e-01}  & \num{9.248000e-01}   & \num{3.247000e-01}  & \num{5.716800e+03}  & \num{1.994600e+03}  & \num{7.879600e+02} \\
$\runid$ & & & \textbf{\textit{MA}} & \num[math-rm=\mathbf]{9.049e-02} & \num[math-rm=\mathbf]{9.5e-02} & \num[math-rm=\mathbf]{2.333000e-01}  & \num[math-rm=\mathbf]{9.431667e-01} & \num[math-rm=\mathbf]{1.616000e-01}  & \num[math-rm=\mathbf]{9.474333e-01} & \num[math-rm=\mathbf]{6.491000e-02}  & \num[math-rm=\mathbf]{9.498000e-01}  & \num[math-rm=\mathbf]{7.134000e-02}  & \num[math-rm=\mathbf]{5.688502e+03}  & \num[math-rm=\mathbf]{1.585400e+03}  & \num[math-rm=\mathbf]{1.137161e+03} \\
\midrule
$\runid$ & \multirow{3}{*}{{\scriptsize \centering\textbf{\textit{C2}}}}
& \multirow{3}{*}{{\scriptsize \centering\textbf{\textit{\#1}}}}
& \textbf{\textit{T}} & \num{1.480e-01} & \num{4.8e-01}  & N/A & N/A & N/A & N/A & \num{2.927000e-01} & \num{5.542000e-01} & \num{4.082000e-01} & N/A & -- & N/A  \\
$\runid$ & & & \textbf{\textit{MP}} & \num{8.855e-02} & \num{1.14e-01}  & \num{3.764000e-01}  & \num{8.791333e-01} & \num{2.554000e-01}  & \num{8.795667e-01} & \num{1.902000e-01}  & \num{7.961000e-01}    & \num{4.697000e-01}  & \num{6.045100e+03}  & \num{2.496000e+03}  & \num{7.488800e+02} \\
$\runid$ & & & \textbf{\textit{MA}} &  \num[math-rm=\mathbf]{9.232e-02}  & \num[math-rm=\mathbf]{7.7e-02} & \num[math-rm=\mathbf]{2.365000e-01}  & \num[math-rm=\mathbf]{9.477667e-01} & \num[math-rm=\mathbf]{1.604000e-01}  & \num[math-rm=\mathbf]{9.480667e-01} & \num[math-rm=\mathbf]{6.072000e-02}  & \num[math-rm=\mathbf]{9.697000e-01}   & \num[math-rm=\mathbf]{6.329000e-02}  & \num[math-rm=\mathbf]{7.892974e+03}  & \num[math-rm=\mathbf]{4.404200e+03}  & \num[math-rm=\mathbf]{9.259420e+02} \\
\bottomrule
\end{tabular}}
\end{scriptsize}
\end{table}
}

\medskip\noindent
\iparagraph{Standalone Tumor Inversion versus Moving Patient versus Moving Atlas}
To compare the different inversion schemes, we show results obtained using the newly introduced \textit{Moving Atlas (MA)} strategy to the results for the \textit{Moving Patient (MP)} solution scheme and those for a tumor standalone \textit{(T)} solver without inter-subject registration, respectively, in~\tabref{tab:ATRV-DIF-CM-3::T-016-dt-001-k-01-sigma-pi-15::MA-vs-MP-L1wL2-vs-L2-simple-obj} for both test cases (STRV-C1) and (STRV-C2).

\medskip\noindent
\iiparagraph{Observations} Solving the single-snapshot tumor inversion problem without inter-subject registration using a tumor standalone solver for some arbitrary normal brain anatomy (atlas) (cf.\,\cite{Gholami:2016a}) exhibits poor reconstruction quality. Our joint registration and biophysical inversion \textit{Moving Atlas} approach  (run $\#3$ and $\#6$ in~\tabref{tab:ATRV-DIF-CM-3::T-016-dt-001-k-01-sigma-pi-15::MA-vs-MP-L1wL2-vs-L2-simple-obj}) results in significantly higher reconstruction quality and outperforms the other strategies (\textit{Moving Patient}, run $\#2$ and $\#5$, and tumor standalone,  run $\#1$ and $\#4$) by a large margin, both in terms of data-misfit and Dice coefficient, as well as in reconstructing the true initial condition.

Comparing against the tumor standalone approach, we improve the reconstruction quality of the input data (grown tumor) from a relative data-misfit of \num{2.90E-1} (run $\#1$) to \num{6.49E-2} (run $\#3$) using the new \textit{Moving Atlas} joint inversion scheme, and likewise improve the reconstruction quality for the tumor initial condition (compared to the ground truth) from a relative $\ell_2-$error of \num{4.25E-1} for tumor standalone (run $\#1$) to \num{7.13E-2} (run $\#3$) for the \textit{Moving Atlas} scheme.
Furthermore, the estimation of the characteristic diffusivity in white matter $k_f$ is significantly improved for the joint inversion scheme (going from a relative error of \num{4.80E-1} (tumor standalone; run $\#4$) to a relative error of \num{7.70E-2} (\textit{Moving Atlas}; run $\#6$)).

\begin{figure}[htb]\centering
  \makebox[\textwidth]{\centering
  \includegraphics[width=1.02\textwidth, trim=5 0 0 0, clip]{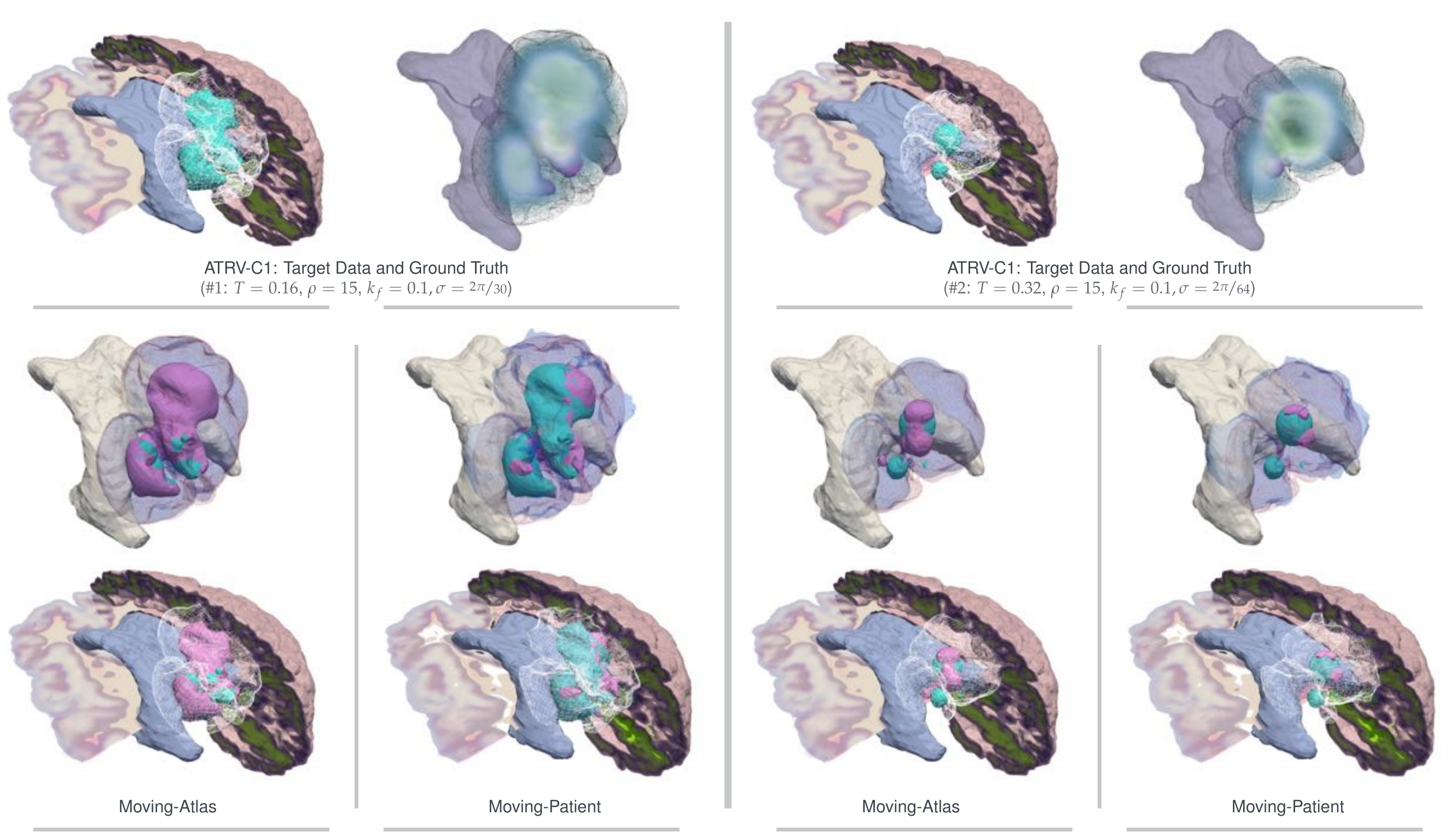}}
  \caption{Comparison of the \ipoint{Moving Patient (MP)}, and the new \ipoint{Moving Atlas (MA)} scheme. We show qualitative results for the (STRV-C1) test case with parameter setting $\#1$ from~\tabref{c8:biophysics:atrv:testcase-setup} (left), and parameter setting $\#2$ from~\tabref{c8:biophysics:atrv:testcase-setup} (right).
  The image shows the reconstructed grown tumor and tumor initial condition for the \ipoint{Moving Atlas} and \ipoint{Moving Patient} solution scheme, featuring different time horizons and sparsity of the initial condition. We show parts of the patient brain anatomy with the respective reconstructed tumor initial condition (magenta wireframe/volume) as compared to the ground truth initial condition (cyan volume). The grown tumor is indicated as white wireframe (3D cut image), and as blue semi-transparent volume compared to the target data given as red wireframe (close-up image).
  The top row shows the test case target data (grown tumor; white wireframe) and initial condition ground truth (cyan volume) for each set of parameters. The close-up image shows a volume rendering of the grown tumor, overlaid with a wireframe indicating the boundary of the tumor resulting from the segmentation (used for visualization purposes). The light blue/white area indicates the ventricles with CSF.
  \label{fig:c8:biophysics:atrv-cm3-ground-truth-T-032-and-T-16-k-01-sigma-pi-32-SIM-TC-large-TOGETHER}
  }
\end{figure}

\begin{SCfigure}[\sidecaptionrelwidth]
  \includegraphics[width=.5\textwidth, trim=0 0 0 0, clip]{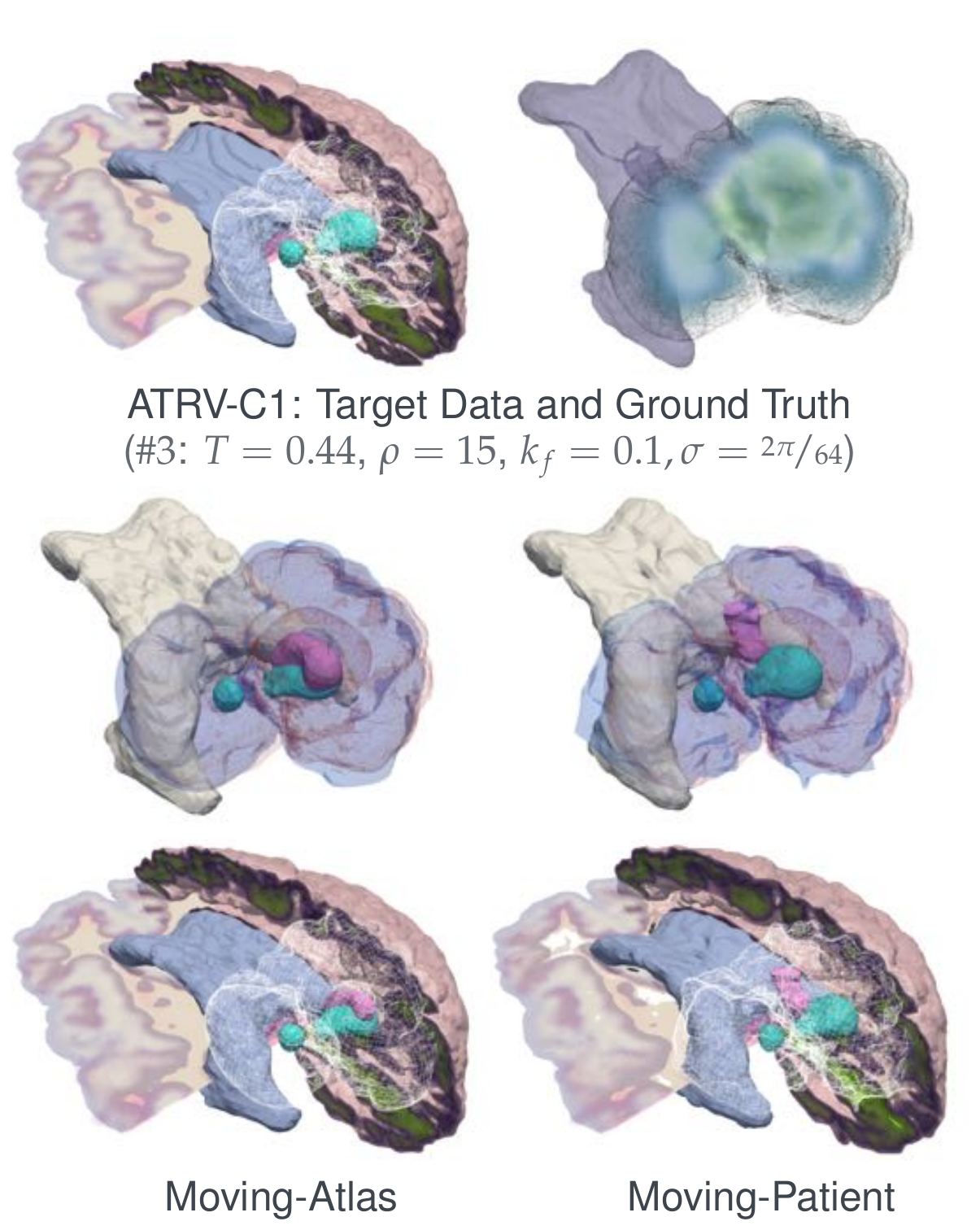}
  \caption{Comparison of the \ipoint{Moving Patient (MP)}, and the new \ipoint{Moving Atlas} scheme. We show qualitative results for the (STRV-C1) test case with parameter setting $\#3$ from~\tabref{c8:biophysics:atrv:testcase-setup}.
  The image shows the reconstructed grown tumor and the tumor initial condition for the \ipoint{Moving Atlas} and \ipoint{Moving Patient} solution scheme. The images show parts of the patient brain anatomy and illustrate the respective reconstructed tumor initial condition (magenta wireframe/volume) as compared to the ground truth initial condition (cyan volume). The grown tumor is indicated as white wireframe (bottom row), and as blue semi-transparent volume compared to the target data given as red wireframe (top row). The light blue/white area indicates the ventricles with CSF.
  The top row shows the test case target data (grown tumor; white wireframe) and initial condition ground truth (cyan volume). The close-up image shows a volume rendering of the grown tumor, overlaid with a wireframe indicating the boundary of the tumor resulting from the segmentation (used for visualization purposes). We observe that the \ipoint{Moving Patient} reconstruction yields a wrong position of the initial condition and inferior reconstruction of the grown tumor.
  \label{fig:c8:biophysics:atrv-cm3-T-044-k-01-sigma-pi-32-MA-vs-MP-L1wL2-vs-L2}
  }
\end{SCfigure}

{
\setlength\tabcolsep{6pt}
\begin{table}[htb]
  \caption{Comparison of the \ipoint{Moving Patient}, and the new \ipoint{Moving Atlas} scheme. We report results for the (STRV-C1) and (STRV-C2) test cases with centers $\vect{x}_{c_1}$ and $\vect{x}_{c_2}$, and parameter setting $\#1$ from~\tabref{c8:biophysics:atrv:testcase-setup}, respectively. Here we are not inverting for $k_f$, but only invert for the tumor initial condition and registration velocity.
  We analyze the sensitivity of the joint inversion with respect to perturbations in the characteristic diffusivity $k_f$ by inverting with different choices of $k_f$; bold numbers correspond to inversion with the ground truth value ($k_f^\star=\num{1E-1}$). We always use the ground truth proliferation rate $\rho_f=15$.
  We report the average mismatch for the anatomy labels for the pathologic brain tissue labels $\elltwoB$, the healthy brain tissue labels $\elltwoHPB$ and the tumor $\elltwoT$, the mean Dice coefficient for brain tissue $\diceB$, for the healthy patient brain tissue $\diceHPB$ and for the tumor $\diceT$ (in cases without Dice score for the tumor reconstruction, the tumor probability map has values below $0.5$ everywhere).
  Furthermore, we measure the $\ell_2$-error $\elltwoINIT$ of the reconstructed tumor initial condition to the ground truth.
  \label{tab:ATRV-DIF-CM-3::T-016-dt-001-k-01-sigma-pi-15::MA-vs-MP-L1wL2-vs-L2-simple-obj-VARY-K}
}
\centering
\begin{scriptsize}
\begin{tabular}{lllr|llllLLL}
\toprule
ID & & & $k_f$ & $\elltwoB$  & $\diceB$ & $\elltwoHPB$  & $\diceHPB$ & $\elltwoT$  & $\diceT$    & $\elltwoINIT$ \\
\midrule
$\runid$ & \multirow{9}{*}{\rotatebox{90}{\scriptsize \centering\textbf{\textit{STRV-C1 (\#1)}}}}
& \multirow{4}{*}{\rotatebox{90}{\scriptsize \centering\textbf{\textit{MP}}}}
& $\num{5E-1}$    & \num{3.799000e-01}  & \num{8.618000e-01} & \num{2.578000e-01}  & \num{8.786333e-01} & \num{3.962000e-01}  & --  & \num{4.993000e-01}  \\
$\runid$ & & &  $\num{3E-1}$  & \num{3.763000e-01}  & \num{8.644333e-01} & \num{2.566000e-01}  & \num{8.789000e-01} & \num{2.878000e-01}  & \num{7.757000e-01}  & \num{8.684000e-01} \\
$\runid$ & & &  $k_f^\star = \num[math-rm=\mathbf]{1E-1}$ & \num[math-rm=\mathbf]{3.716000e-01}  & \num[math-rm=\mathbf]{8.721333e-01} & \num[math-rm=\mathbf]{2.555000e-01}  & \num[math-rm=\mathbf]{8.787667e-01} & \num[math-rm=\mathbf]{1.454000e-01}  & \num[math-rm=\mathbf]{9.209000e-01}   & \num{3.225000e-01}  \\
$\runid$ & & & $\num{1E-3}$  & \num{3.727000e-01}  & \num{8.749333e-01} & \num{2.594000e-01}  & \num{8.729000e-01} & \num{2.491000e-01}  & \num{8.746000e-01}     & \num{5.392000e-01}   \\
\cmidrule{3-11}
$\runid$ & & \multirow{4}{*}{\rotatebox{90}{\scriptsize \centering\textbf{\textit{MA}}}}
& $\num{5E-1}$    & \num{2.396000e-01}  & \num{9.455000e-01} & \num{1.617000e-01}  & \num{9.468667e-01} & \num{5.645000e-01}  & --     & \num{1.372000e+00}  \\
$\runid$ & & &  $\num{3E-1}$ & \num{2.400000e-01}  & \num{9.371667e-01} & \num{1.632000e-01}  & \num{9.477667e-01} & \num{2.863000e-01}  & \num{3.658000e-01}  & \num{9.412000e-01} \\
$\runid$ & & &  $ k_f^\star = \num[math-rm=\mathbf]{1E-1}$ & \num[math-rm=\mathbf]{2.345000e-01}  & \num[math-rm=\mathbf]{9.427667e-01} & \num[math-rm=\mathbf]{1.623000e-01}  & \num[math-rm=\mathbf]{9.475000e-01} & \num[math-rm=\mathbf]{6.773000e-02}  & \num[math-rm=\mathbf]{9.563000e-01}  & \num[math-rm=\mathbf]{8.021000e-02}   \\
$\runid$ & & & $\num{1E-3}$   & \num{2.368000e-01}  & \num{9.481000e-01} & \num{1.607000e-01}  & \num{9.481000e-01} & \num{5.402000e-01}  & --  & \num{7.043000e-01}   \\
\midrule
$\runid$ & \multirow{9}{*}{\rotatebox{90}{\scriptsize \centering\textbf{\textit{STRV-C2 (\#1)}}}}
& \multirow{4}{*}{\rotatebox{90}{\scriptsize \centering\textbf{\textit{MP}}}}
& $\num{3E-1}$   & \num{3.764000e-01}  & \num{8.792667e-01} & \num{2.553000e-01}  & \num{8.799333e-01} & \num{3.657000e-01}  & --   & \num{4.957000e-01}  \\
$\runid$ & & & $\num{2E-1}$  & \num{3.761000e-01}  & \num{8.794667e-01} & \num{2.551000e-01}  & \num{8.801000e-01} & \num{2.638000e-01}  & --  & \num{5.253000e-01}  \\
$\runid$ & & & $k_f^\star = \num[math-rm=\mathbf]{1E-1}$ & \num[math-rm=\mathbf]{3.762000e-01}  & \num[math-rm=\mathbf]{8.796333e-01} & \num[math-rm=\mathbf]{2.552000e-01}  & \num[math-rm=\mathbf]{8.801333e-01} & \num[math-rm=\mathbf]{1.931000e-01}  & \num[math-rm=\mathbf]{8.341000e-01}  & \num[math-rm=\mathbf]{4.240000e-01}  \\
$\runid$ & & & $\num{1E-2}$  & \num{3.823000e-01}  & \num{8.747667e-01} & \num{2.594000e-01}  & \num{8.750333e-01} & \num{2.566000e-01}  & \num{6.857000e-01}  & \num{6.938000e-01}  \\
\cmidrule{3-11}
$\runid$ & & \multirow{4}{*}{\rotatebox{90}{\scriptsize \centering\textbf{\textit{MA}}}}
& $\num{3E-1}$ & \num{2.408000e-01}  & \num{9.459333e-01} & \num{1.629000e-01}  & \num{9.465667e-01} & \num{4.197000e-01}  & --  & \num{5.992000e-01}  \\
$\runid$ & & & $\num{2E-1}$   & \num{2.397000e-01}  & \num{9.462000e-01} & \num{1.624000e-01}  & \num{9.467667e-01} & \num{2.539000e-01}  & --  & \num{7.955000e-01}  \\
$\runid$ & & & $k_f^\star = \num[math-rm=\mathbf]{1E-1}$  & \num[math-rm=\mathbf]{2.365000e-01}  & \num[math-rm=\mathbf]{9.477667e-01} & \num[math-rm=\mathbf]{1.604000e-01}  & \num[math-rm=\mathbf]{9.480667e-01} & \num[math-rm=\mathbf]{6.072000e-02}  & \num[math-rm=\mathbf]{9.697000e-01}  & \num[math-rm=\mathbf]{6.329000e-02}  \\
$\runid$ & & & $\num{1E-2}$  & \num{2.395000e-01}  & \num{9.462333e-01} & \num{1.622000e-01}  & \num{9.468000e-01} & \num{3.390000e-01}  & \num{6.209000e-01} & \num{6.776000e-01}  \\
\bottomrule
\end{tabular}
\end{scriptsize}
\end{table}
}

\begin{figure}[htb]\centering
  \makebox[\textwidth]{\centering
  \includegraphics[width=1.02\textwidth, trim=0 0 0 0, clip]{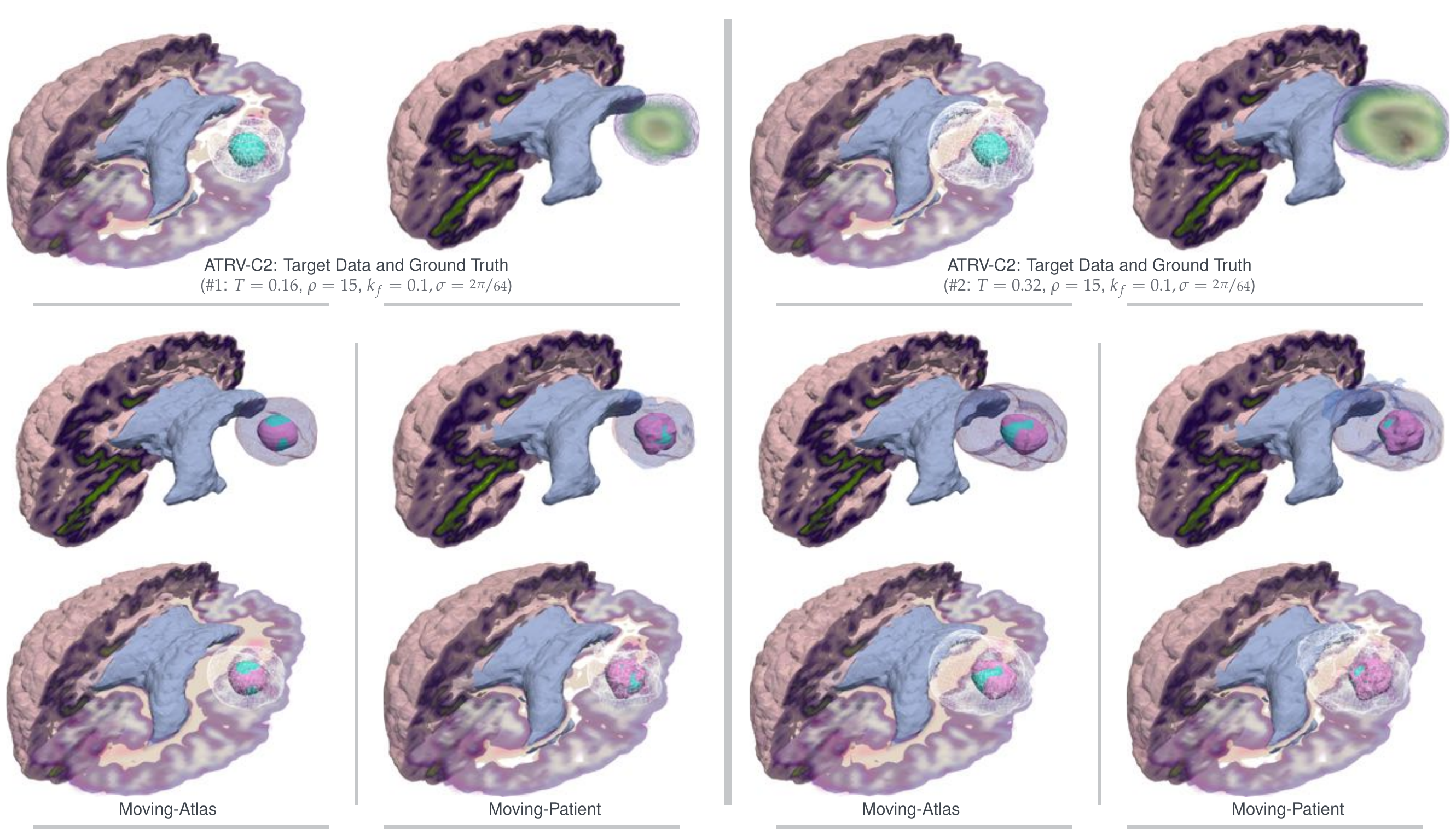}}
  \caption{Comparison of the \ipoint{Moving Patient}, and the new \ipoint{Moving Atlas} scheme. We show qualitative results for the (STRV-C2) test case with parameter setting $\#1$ from~\tabref{c8:biophysics:atrv:testcase-setup} (left image), and parameter setting $\#2$ from~\tabref{c8:biophysics:atrv:testcase-setup} (right image).
  The image shows the reconstructed grown tumor and the tumor initial condition for the \ipoint{Moving Atlas} and \ipoint{Moving Patient} solution scheme featuring different time horizons for the tumor evolution. Illustrated are parts of the (approximated) patient brain anatomy with the respective reconstructed tumor initial condition (magenta wireframe/volume) as compared to the ground truth initial condition (cyan volume). The grown tumor is indicated as white wireframe (bottom images), and as blue semi-transparent volume compared to the target data given as red wireframe (top images).
  The first row shows the test case target data (grown tumor; white wireframe (left) and volume rendering (right)) and initial condition ground truth (cyan volume) for each set of parameters.
  \label{fig:c8:biophysics:atrv-cm5-ground-truth-T-032-and-T-16-k-01-sigma-pi-32-SIM-TC-large-TOGETHER}
  }
\end{figure}

\begin{figure}[h!]\centering
  \emph{Residuals after Reconstruction using the \ipoint{Moving Atlas} and \ipoint{Moving-Patient} Scheme for (STRV-C1)} \\
  \makebox[\textwidth]{\centering
  \includegraphics[width=\textwidth, trim=0 0 0 0, clip]{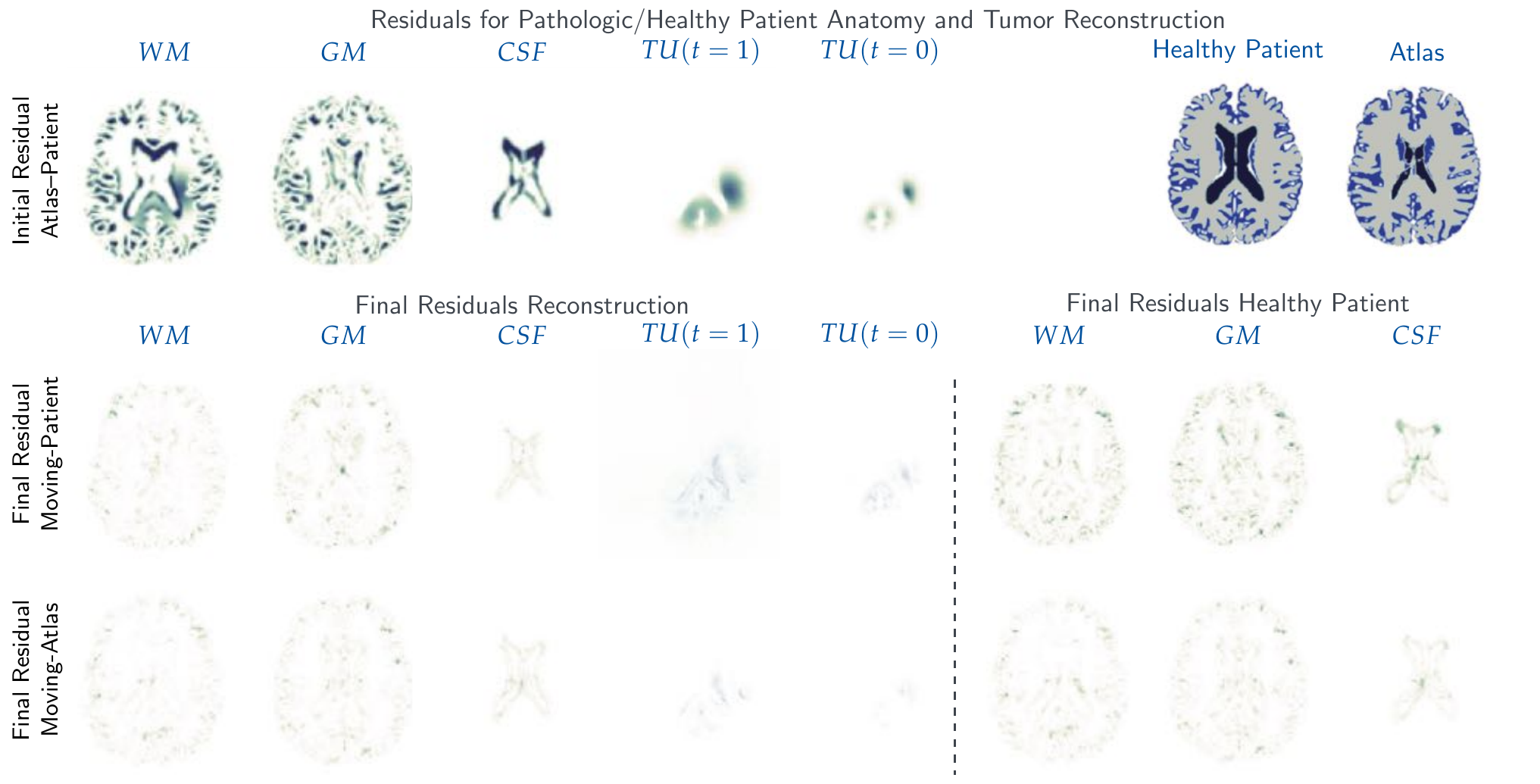}}
  \\ \vspace{5pt}
  \emph{Residuals after Reconstruction using the \ipoint{Moving Atlas} and \ipoint{Moving Patient} Scheme for (STRV-C2)} \\
  \makebox[\textwidth]{\centering
  \includegraphics[width=\textwidth, trim=0 0 0 0, clip]{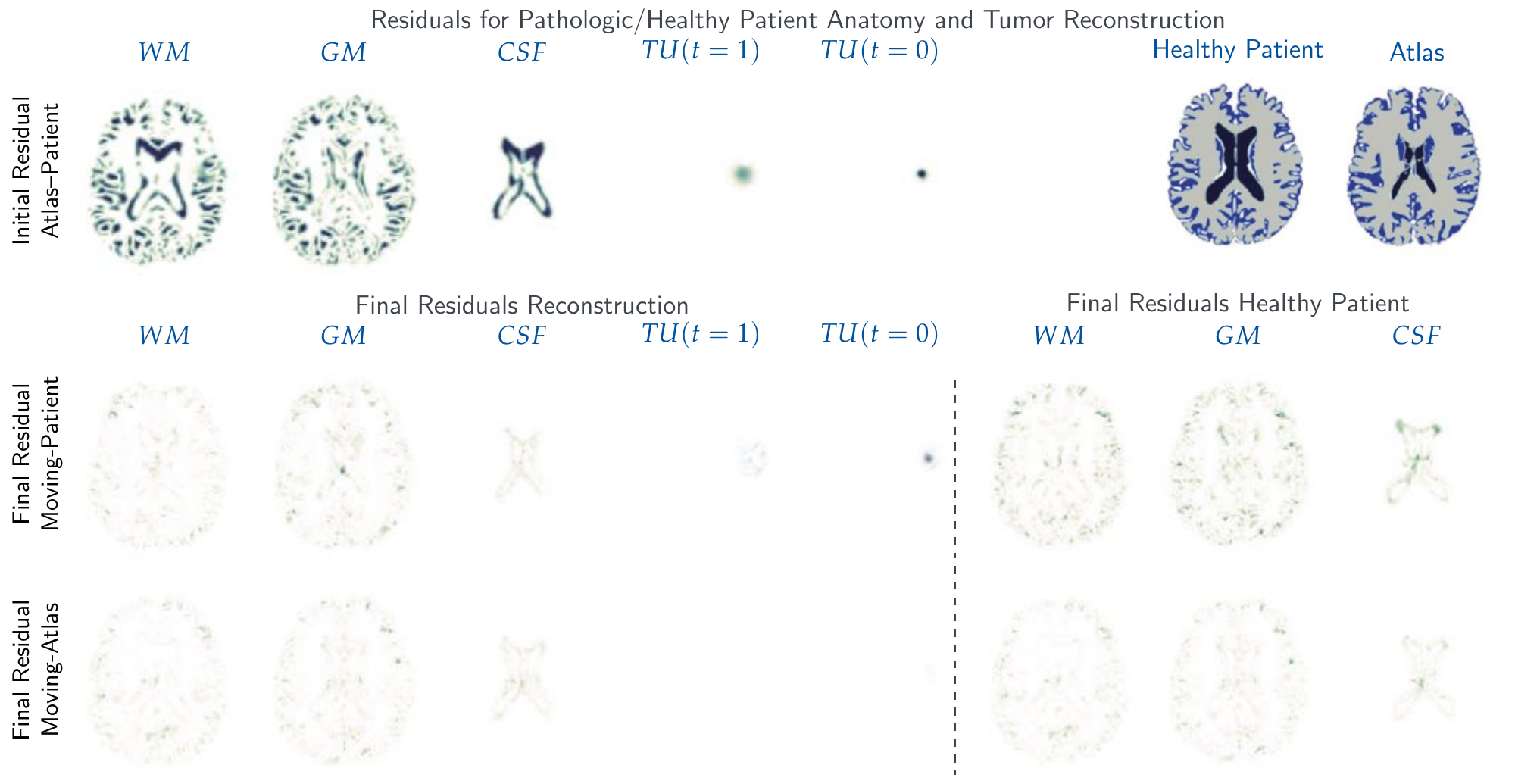}}
  \caption{Comparison of the \ipoint{Moving Patient} Picard iteration scheme, and the new \ipoint{Moving Atlas} Picard iteration scheme. We show  initial and final residuals for the (STRV-C1) test case (top) and the (STRV-C2) test case (bottom) with parameter setting $\#1$ from~\tabref{c8:biophysics:atrv:testcase-setup}, respectively.
  The tableaus show the initial and final residual for the reconstruction of the pathologic and healthy patient for the \ipoint{Moving Atlas} and \ipoint{Moving Patient} solution scheme, respectively for parameter setting $\#1$ from~\tabref{c8:biophysics:atrv:testcase-setup} for both tumor seeds $\vect{x}_{c_1}$ and $\vect{x}_{c_2}$.
  For each case, the top row shows the initial residual between patient input data and the healthy atlas anatomy; the healthy anatomies are illustrated on the right, where gray indicates white matter, blue indicates gray matter, and black indicates cerebrospinal fluid with ventricles.
  The final residual for the reconstruction of the patient input data and the healthy patient anatomy are shown in the second and third row, for joint inversion using the \ipoint{Moving Patient} (run $\#2$ for (STRV-C1) and run $\#5$ for (STRV-C2) in~\tabref{tab:ATRV-DIF-CM-3::T-016-dt-001-k-01-sigma-pi-15::MA-vs-MP-L1wL2-vs-L2-simple-obj}) and the \ipoint{Moving Atlas} scheme (run $\#3$ for (STRV-C1) and run $\#6$ for (STRV-C2) in~\tabref{tab:ATRV-DIF-CM-3::T-016-dt-001-k-01-sigma-pi-15::MA-vs-MP-L1wL2-vs-L2-simple-obj}), respectively. All images show axial slice $64$ (of $128$) of the 3D volume.
  \label{fig:c8:biophysics:atrv-cm3-and-cm5-ground-truth-T-016-k-01-sigma-pi-32_RESIDUALS-opacity}
  }
\end{figure}

\medskip\noindent
\iparagraph{Moving Patient versus Moving Atlas}
For a more detailed comparison of the \textit{Moving Atlas} and the \textit{Moving Patient} scheme, we show qualitative results of the reconstructed tumor initial condition and the grown tumor compared to the target data and ground truth initial condition in~\figref{fig:c8:biophysics:atrv-cm3-ground-truth-T-032-and-T-16-k-01-sigma-pi-32-SIM-TC-large-TOGETHER} (for STRV-C1) and~\figref{fig:c8:biophysics:atrv-cm5-ground-truth-T-032-and-T-16-k-01-sigma-pi-32-SIM-TC-large-TOGETHER} for (STRV-C2). In~\figref{fig:c8:biophysics:atrv-cm3-and-cm5-ground-truth-T-016-k-01-sigma-pi-32_RESIDUALS-opacity}, we outline the initial and final residuals for the different brain tissue labels (white matter (WM), gray matter (GM), cerebrospinal fluid (CSF), grown tumor (TU($t=1$)), and tumor initial condition TU($t=0$); for axial slice 64 of a 3D-volume) before and after the joint registration and biophysical inversion for the test case variants (STRV-C1) and (STRV-C2). 
Furthermore, we display structural differences between the utilized normal brain anatomies for both inversion schemes.
We show qualitative results for a longer time horizon of tumor evolution for the (STRV-C1) test case (with a very sparse initial condition) in~\figref{fig:c8:biophysics:atrv-cm3-T-044-k-01-sigma-pi-32-MA-vs-MP-L1wL2-vs-L2}.

We also study the sensitivity of the \textit{Moving Atlas} and \textit{Moving Patient} solution strategies with respect to perturbations in the tumor model parameters. In~\tabref{tab:ATRV-DIF-CM-3::T-016-dt-001-k-01-sigma-pi-15::MA-vs-MP-L1wL2-vs-L2-simple-obj-VARY-K}, we vary the value of the characteristic diffusivity in white matter from the ground truth, and monitor the obtained reconstruction performance.

\medskip\noindent
\iiparagraph{Observations}
The new \textit{Moving Atlas} scheme outperforms the \textit{Moving Patient} counterpart~\cite{Scheufele:2018} in various ways. \tabref{tab:ATRV-DIF-CM-3::T-016-dt-001-k-01-sigma-pi-15::MA-vs-MP-L1wL2-vs-L2-simple-obj} shows that, for the reconstruction of the brain anatomy (model prediction for anatomy labels of brain tissue labels WM, GM, and CSF), we reach a relative data-misfit ($\ell_2$-error) of \num{2.33E-1} (run $\#3$) compared to \num{3.72E-1} (run $\#2$) for the \textit{Moving Patient} scheme (this translates to a Dice score of \num{9.43E-1} (averaged over all labels) for the \textit{Moving Atlas} compared to a Dice score of \num{8.72E-1} for the old scheme). Similarly, for the approximation of the actual healthy patient anatomy, we improve from a Dice score of \num{8.79E-1} (relative $\ell_2$-error of $25.6\%$; run $\#3$) obtained from the \textit{Moving Patient} solution to a Dice score of \num{9.47E-1} (relative $\ell_2$-error of only $16.2\%$; run $\#2$) using the new scheme. \figref{fig:c8:biophysics:atrv-cm3-and-cm5-ground-truth-T-016-k-01-sigma-pi-32_RESIDUALS-opacity} illustrates the structural differences between the atlas and patient brain, the initial residuals and the residuals after joint inversion using the \textit{Moving Atlas} and the \textit{Moving Patient} scheme, respectively, for an exemplary slice (axial slice 64) of the 3D volume. We observe smaller errors for the \textit{Moving Atlas} solution. The improved anatomy reconstruction is characteristic for the \textit{Moving Atlas} strategy; we observe similar trends for various solver and parameter configurations (compare also~\tabref{tab:ATRV-DIF-CM-3::T-016-dt-001-k-01-sigma-pi-15::MA-vs-MP-L1wL2-vs-L2-simple-obj-VARY-K} with inversion under perturbed or wrong tumor model parameters).

The \textit{Moving Atlas} scheme furthermore results in improved reconstruction quality of the predicted grown tumor and higher similarity to the target data. For (STRV-C1), the \textit{Moving Atlas} scheme results in a relative error of $6.4\%$ (Dice score of \num{9.50E-1}; run $\#3$) compared to a relative error of $14.4\%$ (Dice score of \num{9.25E-1}; run $\#2$) for the \textit{Moving Patient} solution. The numbers for (STRV-C2) are very similar with $6.0\%$ relative error (Dice score of \num{9.70E-1}; run $\#6$) versus $19.0\%$ relative error (Dice score of \num{7.96E-1}; run $\#5$).

The primary objective for the derivation of the \textit{Moving Atlas} scheme was to allow for a more informative inversion for biophysical model parameters. We primarily consider the estimation of the tumor initial condition, but also invert for the characteristic diffusivity $k_f$ in white matter. The relative errors\footnote{Note, that for the \textit{Moving Atlas} scheme, the reconstructed initial condition naturally ``lives'' in the patient space; for the \textit{Moving Patient} solution strategy, we invert for the tumor initial condition in the atlas space. For a fair comparison, the reconstructed initial condition for the \textit{Moving Patient} scheme is advected to the patient space before computing the relative error to the ground truth.} $\elltwoINIT$ for the inversion of the ground truth tumor initial condition in~\tabref{tab:ATRV-DIF-CM-3::T-016-dt-001-k-01-sigma-pi-15::MA-vs-MP-L1wL2-vs-L2-simple-obj} obtained for each scheme clearly attest the \textit{Moving Atlas} scheme to be more reliable and sound in recovering biophysical parameters\footnote{Our analysis assumes that tumor-growth is perfectly described by a reaction-diffusion model, which is quite certainly not true. When we say ``biophysically meaningful'' this is to be seen under the aforementioned assumption. Note furhermore, that the tumor model can easily be exchanged in the modular setting of our Picard iteration approach.}.
Using the improved \textit{Moving Atlas} scheme, we achieve excellent reconstruction of the true initial condition with a relative error of only $7.1\%$ (for (STRV-C1); run $\#3$) and $6.3\%$ (for (STRV-C2); run $\#6$), as opposed to a relative error of $32.5\%$ (for (STRV-C1); run $\#2$) and $42.4\%$ (for (STRV-C2); run $\#5$), respectively, when using the \textit{Moving Patient} scheme instead. An illustration of the error is given in~\figref{fig:c8:biophysics:atrv-cm3-and-cm5-ground-truth-T-016-k-01-sigma-pi-32_RESIDUALS-opacity} (for an axial cut at slice 64).
We explain the improved inversion properties of the \textit{Moving Atlas} scheme by its general idea to seek for a good approximation of the healthy patient brain anatomy first, and, thus, to carry out the inversion in the ``correct'' space. For the \textit{Moving Patient} scheme, the inversion in the ``wrong'' anatomy (atlas space) has the potential to induce large errors for the estimation of biophysical parameters as shown in~\secref{sec_scheuf:results_Artificial_Brain} since the inter-subject deformation map can aid matching the pathologic brains. In particular, there is no implication that this deformation map produces reasonable results when applied to the tumor initial condition (or other quantities that describe the time point of tumor genesis) in order to translate the latter into the individual patient application space.
Visual inspection of our simulation results for longer time horizons in~\figref{fig:c8:biophysics:atrv-cm3-T-044-k-01-sigma-pi-32-MA-vs-MP-L1wL2-vs-L2} support this point: The \textit{Moving Patient} solution recovers a wrong position of the tumor initial condition (compare the green (ground truth) and purple (reconstruction) iso-volume representations of the initial tumor condition.
A wrong position, wrong shape, or wrong sparsity of the initial condition can, in a second step, also cause a wrong estimation of the characteristic proliferation rate or tumor cell infiltration rate. For our experiments, the difference for the estimation of the characteristic diffusivity $k_f$ is, however, not significant; both schemes yield comparable results. One possible reason for this could be that the diffusion part of the gradient initially is large but then flattens out compared to the gradient component for the initial condition parametrization. The theoretical and numerical analysis of this effect is ongoing work.

\section{Conclusion}
\label{sec_scheuf:conclusion}
\medskip\noindent
We present a new inverse problem formulation to solve the patient-specific tumor inversion problem given a single data-snapshot in time only. We remedy the lack of healthy patient imaging data via registration to a healthy atlas brain, resulting in a joint inter-subject registration and biophysical inversion scheme. We simultaneously solve for an estimation of the inter-subject deformation map, and (a subset of) the biophysical model parameters for tumor progression. More specifically, we solve for an estimation of a sparse tumor initial condition, and for the characteristic diffusivity rate of tumor cells in white matter.
The presented scheme conceptually improves on our earlier coupled solution approach~\cite{Scheufele:2018}. The latter uses the registration to warp patient target data towards the atlas space, and inverts for tumor model parameters in the atlas brain (proxy for the unknown healthy patient brain). This may introduce large errors due to unwanted data fitting and model inversion in a wrong anatomy. We demonstrate this for synthetic cases. The new scheme computes an approximation of the healthy patient and performs the tumor inversion in the partient space. We furthermore encourage a sparse localization of the initial tumor (presented in our work~\cite{Subramanian19aL1}) to allow for a more reliable reconstruction of the model parameters. This biophysically motivated constraint can help to understand tumor genesis and its role in the subsequent tumor evolution.
We derive a Picard-iteration-type solution strategy by dividing the strongly coupled set of first order optimality conditions into two solvers.

For our new method, we observe  \bipa \item improved similarity of the reconstructed probability maps of brain tissue labels (WM, GM, CSF) with respect to the patient data, \item significantly smaller errors for the reconstruction of the grown tumor, \item a better approximation of the actual healthy patient brain anatomy, and, \item more reliable reconstruction of biophysical parameters, such as the tumor initial condition. \eipa

The reconstruction result obtained from the \textit{Moving Atlas} scheme seems to be more ``natural'' as we invert in the ``correct'' brain anatomy of the (healthy) patient, and the registration does not act on the tumor probability map (as is the case for the \textit{Moving Patient} strategy).
In particular, using the \textit{Moving Atlas} scheme, we were able to improve the Dice score for the brain anatomy from \num{8.72E-1} for the \textit{Moving Patient} strategy to \num{9.43E-1}; the Dice score for the approximation of the healthy patient brain anatomy from \num{8.79E-1} for \textit{Moving Patient} to \num{9.47E-1}; and the Dice coefficient for the tumor reconstruction from \num{8.34E-1} for \textit{Moving Patient} to \num{9.70E-1}.
We observe an excellent reconstruction of the true initial condition with a relative error of only $7.1\%$ (STRV-C1) and $6.3\%$ (STRV-C2) (compared to $32.3\%$ and $42.4\%$ for \textit{Moving Patient}) using the more sophisticated \textit{Moving Atlas} scheme. We further conclude, that a sparse localization of the initial condition is essential if targeting biophysical parameter estimation from patient MRI. The previously used $\ell_2$-regularization results in rich initial conditions.
Lastly, the \textit{Moving Atlas} solution scheme seems to be more useful in identifying wrong parameters (such as cell proliferation rate and cell migration rate of brain tissue) than the \textit{Moving Patient} counterpart.

\FloatBarrier

\section*{Acknowledgements}

This material is based upon work supported by a DAAD (German Academic Exchange Service) scholarship granted to the first author and the Tinsley Oden Fellowship, granted to the last author. The work is further supported by AFOSR grant FA9550-17-1-0190; by Simons Foundation grant (Award Number 586055, AM); by NSF grant CCF-1337393; by the U.S. Department of Energy, Office of Science, Office of Advanced Scientific Computing Research, Applied Mathematics program under Award Numbers DE-SC0010518 and DE-SC0009286; by NIH grant 10042242; by DARPA grant W911NF-115-2-0121; and by the University of Stuttgart, Institute for Parallel and Distributed Systems. Any opinions, findings, and conclusions or recommendations expressed herein are those of the authors and do not necessarily reflect the views of the AFOSR, DOE, NIH, DARPA, and NSF. Computing time on the High-Performance Computing Centers (HLRS) Hazel Hen system (Stuttgart, Germany) was provided by an allocation of the federal project application ACID-44104.

\vspace{-3mm}
\bibliography{literature}
\bibliographystyle{siam}
\end{document}